\documentclass[twocolumn,trackchanges]{aastex62}

\usepackage{amsmath}
\usepackage{multirow}
\usepackage{color}

\newcommand{\hii}{\ion{H}{2}}
\newcommand{\nh}{$n_{\rm H}$}
\newcommand{\gof}{$G_{\rm 0}$}
\newcommand{\goob}{$\langle G_{0} \rangle^{\rm OB}$}
\newcommand{\goisrf}{$\langle G_{0} \rangle^{\rm ISRF}$}

\newcommand{\cii}{\mbox{[\ion{C}{2}]}}

\newcommand{\ois}{\mbox{[\ion{O}{1}]}}
\newcommand{\oi}{\mbox{[\ion{O}{1}]\,63\,$\mu$m}}

\newcommand{\oii}{\mbox{[\ion{O}{1}]\,145\,$\mu$m}}

\newcommand{\oiii}{\mbox{[\ion{O}{3}]\,88\,$\mu$m}}

\newcommand{\oiiii}{\mbox{[\ion{O}{3}]\,52\,$\mu$m}}
\newcommand{\nit}{[\ion{N}{2}]}
\newcommand{\nii}{[\ion{N}{2}]\,122\,$\mu$m}

\newcommand{\niii}{[\ion{N}{3}]\,57\,$\mu$m}

\newcommand{\NH}{$N_{\rm H}$}

\newcommand{\sfir}{$\Sigma_{\rm FIR}$}
\newcommand{\firmol}{$L_{\rm FIR}/M_{\rm mol}$}

\newcommand{\sratio}{$S_{63\,\mu\mathrm{m}}/S_{122\,\mu\mathrm{m}}$}

\shorttitle{SHINING, A Survey of Far Infrared Lines in Nearby Galaxies}
\shortauthors{Herrera-Camus et al.}

\begin{document}

\title{SHINING, A Survey of Far Infrared Lines in Nearby Galaxies. II: Line-Deficit Models, AGN impact, [CII]-SFR Scaling Relations, and Mass-Metallicity Relation in (U)LIRGS}

\correspondingauthor{R. Herrera-Camus}
\email{rhc@mpe.mpg.de}

\author{R. Herrera-Camus}
\affil{Max-Planck-Institut f\"{u}r Extraterrestrische Physik (MPE), Giessenbachstr., D-85748 Garching, Germany}

\author{E. Sturm}
\affiliation{Max-Planck-Institut f\"{u}r Extraterrestrische Physik (MPE), Giessenbachstr., D-85748 Garching, Germany}

\author{J. Graci\'a-Carpio}
\affiliation{Max-Planck-Institut f\"{u}r Extraterrestrische Physik (MPE), Giessenbachstr., D-85748 Garching, Germany}

\author{D. Lutz}
\affiliation{Max-Planck-Institut f\"{u}r Extraterrestrische Physik (MPE), Giessenbachstr., D-85748 Garching, Germany}

\author{A. Contursi}
\affiliation{Max-Planck-Institut f\"{u}r Extraterrestrische Physik (MPE), Giessenbachstr., D-85748 Garching, Germany}

\author{S. Veilleux}
\affiliation{Department of Astronomy and Joint Space-Science Institute, University of Maryland, College Park, MD 20742, USA}

\author{J. Fischer}
\affiliation{Naval Research Laboratory, Remote Sensing Division, 4555 Overlook Avenue SW, Washington DC 20375, USA}

\author{E. Gonz\'alez-Alfonso}
\affiliation{Departamento de F\'isica y Matem\'aticas, Universidad de Alcal\'a, Campus Universitario, E-28871 Alcal\'a de Henares, Madrid, Spain}

\author{A. Poglitsch}
\affiliation{Max-Planck-Institut f\"{u}r Extraterrestrische Physik (MPE), Giessenbachstr., D-85748 Garching, Germany}

\author{L. Tacconi}
\affiliation{Max-Planck-Institut f\"{u}r Extraterrestrische Physik (MPE), Giessenbachstr., D-85748 Garching, Germany}

\author{R. Genzel}
\affiliation{Max-Planck-Institut f\"{u}r Extraterrestrische Physik (MPE), Giessenbachstr., D-85748 Garching, Germany}

\author{R. Maiolino}
\affiliation{Kavli Institute for Cosmology, University of Cambridge, Madingley Road, Cambridge CB3 0HA, UK}

\author{A. Sternberg}
\affiliation{Raymond and Beverly Sackler School of Physics \& Astronomy, Tel Aviv University, Ramat Aviv 69978, Israel}

\author{R. Davies}
\affiliation{Max-Planck-Institut f\"{u}r Extraterrestrische Physik (MPE), Giessenbachstr., D-85748 Garching, Germany}

\author{A. Verma}
\affiliation{Oxford University, Dept. of Astrophysics, Oxford OX1 3RH, UK}



\begin{abstract}
The SHINING survey \citep[Paper~I; ][]{rhc_rhc18a} offers a great opportunity to study the properties of the ionized and neutral media of galaxies from prototypical starbursts and active galactic nuclei (AGN) to heavily obscured objects. Based on {\it Herschel}/PACS observations of the main far-infrared (FIR) fine-structure lines, in this paper we analyze the physical mechanisms behind the observed line deficits in galaxies, the apparent offset of luminous infrared galaxies (LIRGs) from the mass-metallicity relation, and the scaling relations between \cii~158~$\mu$m line emission and star formation rate (SFR). Based on a toy model and the {\it Cloudy} code, we conclude that the increase in the ionization parameter with FIR surface brightness can explain the observed decrease in the line-to-FIR continuum ratio of galaxies. In the case of the \cii\ line, the increase in the ionization parameter is accompanied by a reduction in the photoelectric heating efficiency and the inability of the line to track the increase in the FUV radiation field as galaxies become more compact and luminous. In the central $\sim$kiloparsec regions of AGN galaxies we observe a significant increase in the \oi/\cii\ line ratio; the AGN impact on the line-to-FIR ratios fades on global scales. Based on extinction-insensitive metallicity measurements of LIRGs we confirm that they lie below the mass-metallicity relation, but the offset is smaller than those reported in studies that use optical-based metal abundances. Finally, we present scaling relations between \cii\ emission and SFR in the context of the main-sequence of star-forming galaxies.
\end{abstract}




\section{Introduction} \label{sec:intro}

The far-infrared (FIR) fine-structure lines of C, N, and O offer a powerful tool to characterize the interstellar medium (ISM) of nearby and high-$z$ galaxies. In this context, the SHINING survey of galaxies (``Survey with Herschel of the ISM in Nearby INfrared Galaxies''; PI Sturm) was planned with the purpose of obtaining a comprehensive view of the physical processes at work in the ISM of galaxies, ranging from moderately star-forming to the most dense and obscured environments in luminous infrared galaxies (LIRGs) and around active galactic nuclei (AGN). For this, we used {\it Herschel} PACS to observe the six main FIR atomic and ionic lines in the $\sim55-200~\mu$m range. The description of the survey and the main general results are presented in Paper I \citep{rhc_rhc18a}. 

One of the open questions in the study of the ISM of galaxies is what are the physical mechanisms that drive the decrease in the line-to-FIR continuum ratios --commonly referred to as ``line deficit''-- in the most dense, energetic galactic environments. In the case of the brightest of the FIR fine-structure lines, the ionized carbon line \cii\ at 157.74~$\mu$m, 
the \cii/FIR continuum ratio typically range from $\sim10^{-2}$ in normal, star-forming galaxies to $\sim10^{-4}$ in galaxies with a buried, compact, luminous nucleus \citep[e.g.,][]{rhc_malhotra97,rhc_malhotra01,rhc_luhman98,rhc_luhman03,rhc_gracia-carpio11,rhc_farrah13,rhc_diaz-santos13,rhc_ibar15,rhc_gonzalez-alfonso15,rhc_smith17,rhc_contursi17,rhc_rhc18a}. 

Several physical effects have been suggested to explain the \cii\ deficit, including the reduction of the photoelectric heating efficiency due to the charging or destruction of the small dust grains \citep{rhc_malhotra01,rhc_croxall12}; high-ionization parameters in ionized regions that cause an increase in the fraction of UV photons absorbed by dust relative to the UV photons available to ionized and excite the neutral gas \citep{rhc_luhman03,rhc_abel09,rhc_gracia-carpio11,rhc_fischer14}; the impact of active galactic nuclei (AGN) on the ionization state of the gas \citep{rhc_sargsyan12,rhc_langer15}; strong continuum extinction at 158~$\mu$m; the saturation of the upper energy level at gas temperatures higher than the excitation temperature of the line \citep{rhc_stacey10,rhc_munoz15}; dense PDRs with gas densities higher than the critical density of the \cii\ line, among others. One of the main goals of this paper is to use the sophisticated theoretical models available \citep[e.g, Cloudy, PDR Toolbox; ][]{rhc_ferland13,rhc_fischer14,rhc_kaufman06,rhc_pound08} to interpretat the line-to-FIR continuum trends and understand what are the main physical processes behind the observed line deficits in galaxies.

The detailed characterization of the FIR fine-structure lines in nearby galaxies is also very relevant in the current era of sensitive interferometers such as ALMA and NOEMA. These observatories have made possible detections and spatially resolved observations of galaxies from the epoch of the peak of cosmic star formation to the era of reionization in the rest-frame FIR lines. \cii~158~$\mu$m \citep[e.g.,][]{rhc_debreuck14,rhc_pentericci16,rhc_riechers14,rhc_capak15,rhc_inoue16}, \nii\ \citep[e.g.,][]{rhc_ferkinhoff15,rhc_pavesi16}, and \oiii\ \citep[e.g.,][]{rhc_nagao12,rhc_carniani17}, have been detected in galaxies ranging from typical, star-forming galaxies to extreme starbursts. In this paper we present scaling relations between the \cii\ luminosity and the star formation rate (SFR) as a function of various galaxy properties (separation from the main-sequence of galaxies, star formation efficiency, FIR surface brightness, galaxy type), which can be useful for the interpretation of the growing number of high-$z$ galaxies detected in \cii\ line emission.

Finally, another strong asset of the FIR lines is their ability to penetrate extremely high dust column densities and characterize the conditions in the most obscured objects\footnote{Note, however, that in extreme dust environments, such as that found in Arp~220, the dust optical depth is $\gtrsim1$ in the far-infrared wavelength range up to $\sim850~\mu$m \citep[e.g.,][]
{rhc_sakamoto08,rhc_rangwala11}}. 
One immediate application is to use the FIR lines to determine the metal abundance of LIRGs \citep[e.g.,][]{rhc_nagao11,rhc_fo16,rhc_ps17}. LIRGs are believed to lie below the well know mass-metallicity relation for star-forming galaxies \citep[e.g.,][]{rhc_tremonti04,rhc_rupke08,rhc_kilerci14}, although these results are based on optical-based metallicity measurements. One possibility is that the gas in LIRGs has indeed lower metal content compared to normal galaxies with similar stellar masses. The reason is the accretion of low metallicity gas --triggered by their history of interactions-- from the outskirts towards the central regions \citep[][]{rhc_rupke10,rhc_torrey12}. The other alternative is that the optical-based metallicity measurements in LIRGs are underestimated due to strong dust extinction. The spectral coverage and unprecedent sensitivity offered by {\it Herschel} provide us now with the opportunity to use FIR lines as extinction-free metallicity tracers and to test which of the two scenarios described above is responsible for the observed offset of LIRGs from the mass-metallicity relation. 


This paper is organized as follows. In Section~\ref{sample} we give a brief introduction to the SHINING sample of galaxies. In Section~\ref{sec:analysis} we analyze the observed line-to-continuum trends using a toy model (\S\ref{toy}) and the {\it Cloudy} code (\S\ref{cloudy}). We investigate the influence of AGN emission on the \cii\ and \oi\ line-to-FIR ratios in \S\ref{section:AGN}. We present scaling relations between the \cii\ line emission and the star formation rate in \S\ref{CIISFR_scaling}. Finally, we revisit the observed mass-metallicity relation in (U)LIRGs, this time using the \oiii/\niii\ line ratio as an extinction-insensitive metallicity tracer  (\S\ref{M-Z}). We present our summary and conclusions in Section~\ref{conclusions}.

\section{Galaxy sample}\label{sample}

A detailed description of the SHINING sample characteristics, observational strategy, data reduction, and flux measurements can be found in Paper I \citep[][]{rhc_rhc18a}. Here we provide a brief summary of these topics. 

The SHINING sample consists of 52 nearby  ($z < 0.2$) galaxies that were observed with the PACS spectrometer \citep{rhc_poglitisch10} on board {\it Herschel} \citep{rhc_pilbratt10}. The breakdown by galaxy type is: 8 star-forming, 23 AGNs, and 21 (U)LIRGs. SHINING include some archetypical galaxies such as M~82, M~83, NGC~253, NGC~1068, Circinus, and Arp~220.

The survey includes observations of the six ionized and PDR lines in the $55-200~\mu$m range. These are:  \cii~158~$\mu$m, \oii, \nii, \oiii, \oi, and \niii. 
The data were reduced using HIPE v13.0 (Herschel Interactive Processing Environment\footnote{HIPE is a joint development by the Herschel Science Ground Segment Consortium, consisting of ESA, the NASA Herschel Science Center, and the HIFI, PACS and SPIRE consortia.}; \citealt{rhc_ott10}). In the case of galaxies that were considered point sources for {\it Herschel} (the PACS spectrometer point spread function is $\sim$6\arcsec--11\arcsec\ in the instrument wavelength range) we applied a point-source correction determined both theoretically and from dedicated PV observations \citep{rhc_poglitisch10}. The integrated line and continuum fluxes for all the SHINING sources are listed in Paper I \citep[Table~6 and 7; ][]{rhc_rhc18a}.

In addition to the SHINING galaxies, in this paper we include a compilation of previous ISO extragalactic observations taken from the literature \citep{rhc_malhotra01,rhc_negishi01,rhc_luhman03,rhc_lutz03,rhc_brauher08}, and PACS \cii\ observations of local starburst, (U)LIRGs and AGN by \cite{rhc_sargsyan12} and \cite{rhc_farrah13}. The FIR sizes for the SHINING and the ancillary galaxies were drawn from \cite{rhc_lutz16} and were derived from a 2-dimensional Gaussian fit to the 70~$\mu$m image of the galaxy, with PSF width subtracted in quadrature. We also include in our analysis high-$z$ ($z>4$) galaxies detected in \cii\ line and continuum emission and with size measurements available \citep{rhc_walter12,rhc_riechers13,rhc_debreuck14,rhc_capak15}.

Throughout the paper far-infrared luminosities were measured using the definition given in \citet{rhc_helou88}:

\begin{equation}\label{eq:fir}
\begin{split}
F_{\rm FIR}(42.5\,\mu\rm{m}-122.5\,\mu\rm{m}) = 1.26 \times 10^{-14} \\ 
\times(2.58\ S_{60\,\mu\rm{m}} + S_{100\,\mu\mathrm{m}}),
\end{split}
\end{equation}

\section{Analysis}\label{sec:analysis}

In Paper~I we show how the line-to-FIR ratios of galaxies decrease as a function of FIR luminosity, star-formation efficiency, and FIR surface brightness. In the case of the \cii\ and \ois\ lines, the dispersion in the relation between the line-to-FIR ratio and \sfir\ is only $\sim0.3$~dex over almost five orders of magnitude in \sfir\ \citep[see also][]{rhc_lutz16,rhc_diaz-santos17}. With the purpose of understanding the existence and tightness of these correlations, in this section we present an analysis based on a {\it toy model} powered by the PDR toolbox (\S3.1) and the {\it Cloudy} code (\S3.2).

\subsection{A toy model to explore the PDR lines deficit as a function of $\Sigma_{\rm FIR}$}\label{toy}

We construct a {\it toy model} with the aim of exploring what drives the observed decline of the ratio between the PDR lines (\cii, and \ois~63 and 145~$\mu$m) and the FIR continuum emission as a function of \sfir. Based on the work by \cite{rhc_wolfire90}, we consider two ``extreme'' scenarios for the distribution of stars and gas clouds in galaxies. These scenarios set characteristic values for the physical conditions in the ISM of the galaxy including the density of the neutral gas clouds (\nh), the FUV radiation field intensity impinging upon (\gof), and the FIR continuum and line intensity they emit. The goal is to explore what is driving the line deficits as galaxies become more compact and/or luminous. 

\subsubsection{Scenario 1: Dense PDR}

In this scenario we assume that OB stars are closely associated with molecular clouds. In this case the incident FUV flux on the molecular clouds is dominated by the nearest stars, and \gof\ can be approximated as:

\begin{equation}
\langle G_{0} \rangle^{\rm OB}  \simeq 10^{-2}\Bigg(\frac{L_{*}}{L_{\odot}}\Bigg)\Bigg(\frac{r}{1~{\rm pc}}\Bigg)^{-2},
\end{equation}

\noindent where $L_{*}$ is the average stellar or OB association FUV luminosity, and $r$ is the distance from the stars to the molecular cloud.

\begin{figure*}[t!]
\begin{center}
\includegraphics[scale=0.17]{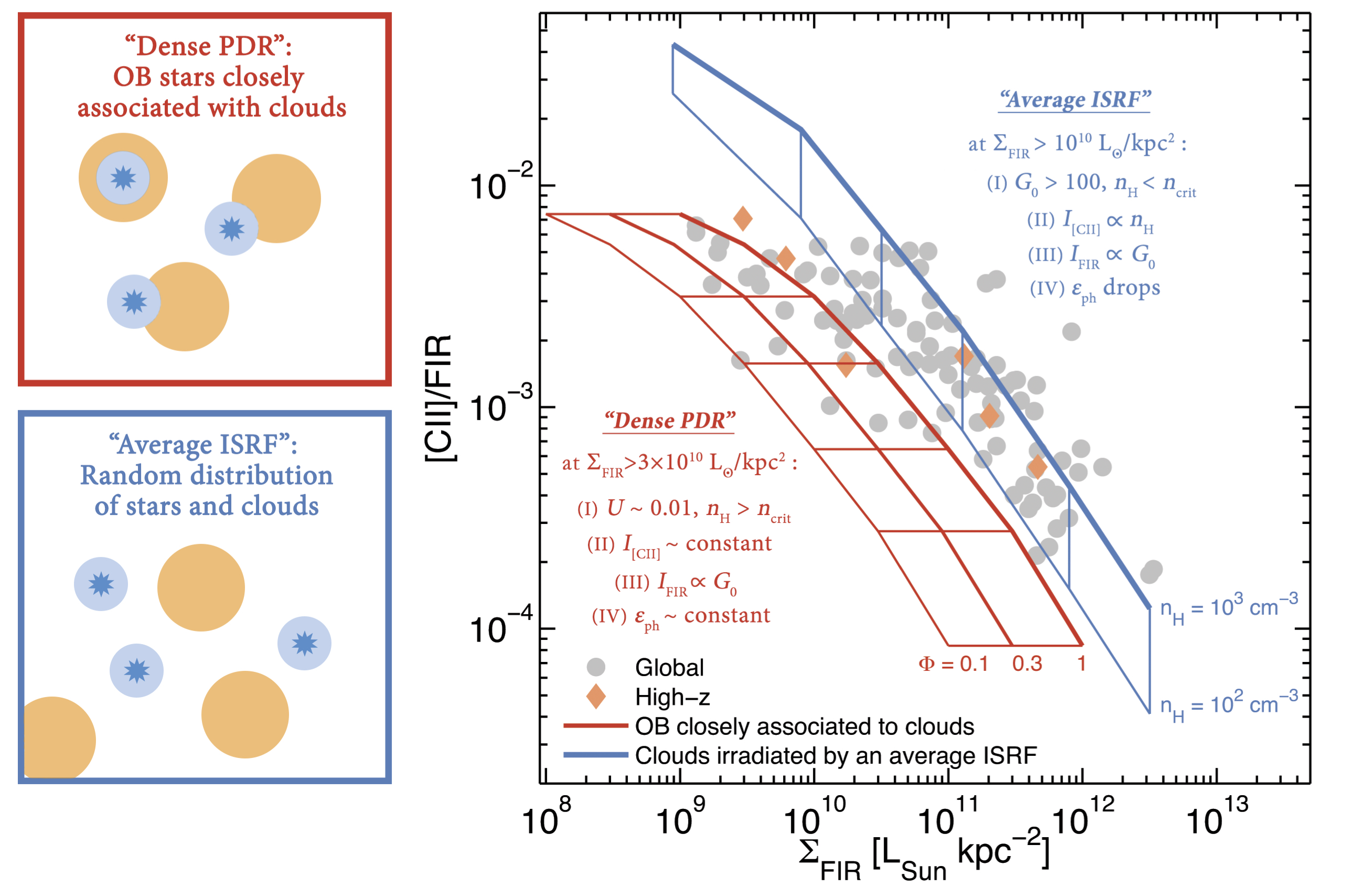}
\caption{\cii-to-FIR ratio as a function of \sfir\ observed in nearby (grey circles) and high-$z$ ($z>4$) galaxies (orange diamonds). To explore the physical processes that drive the \cii\ deficit we consider a {\it toy model} defined by two extreme scenarios; one where all OB stars are closely associated with molecular gas clouds (red box), and one where there OB associations and gas clouds are randomly distributed (blue box). We overplot the model results for both scenarios: (1) dense PDRs varying the beam area filling factor $\Phi$ from 0.1 to 1 (red lines), and (2) random distribution of OB stars and clouds varying the neutral gas density \nh\ from 10$^2$~cm$^{-3}$ to 10$^3$~cm$^{-3}$ (blue lines). 
\label{Fig:cii}}
\end{center}
\end{figure*} 

Consider now the idealized problem of a static, spherically symmetric equilibrium \hii\ region ionized by a point source as discussed by \cite{rhc_draine11}. If the product of the rate of ionizing photons ($Q_0$) and the rms density of the ionized gas ($n_{\rm rms}$) is $(Q_{0}/10^{49}~{\rm s}^{-1})n_{\rm rms} \gg 10^{2}~{\rm cm}^{-3}$, then the radiation pressure acts to concentrate the gas in a spherical shell. This implies that the pressure of ionized gas at the edge of the \hii\ region is higher than obtained in the uniform-density \hii\ region approximation. Conversely, the external pressure confining the \hii\ region has to be higher.

For our calculation of the PDR line-to-continuum ratio we start with a stellar cluster with an ionizing photon rate $Q_{\rm 0}=10^{52}~{\rm s}^{-1}$, which is typical for giant \hii\ regions such as 30~Dor \citep{rhc_kennicutt91}. 
Based on $Q_{\rm 0}$ we estimate the stellar cluster FUV luminosity following $L_{\rm FUV} = 5.3\times10^{5}(Q_{\rm 0}/10^{49}~s)~L_{\odot}$ \citep{rhc_kaufman06}. Then, for a given \hii\ region radius $R$ we determine $p_{\rm edge}$ using the \cite{rhc_draine11} model calculations \citep[e.g., see the model grid results in Figure~11 of][]{rhc_draine11}. We use a standard model that adopts values of $\beta=3$ and $\gamma=10$, where $\beta$ is the ratio of the power in non-ionizing photons to the power in photons with h$\nu>13.6$~eV, and $\gamma$ is a dimensionless parameter that depends on the gas temperature and the mean ionizing photon energy. 

We then solve for \nh\ by equating the ionized gas pressure at the edge of the \hii\ region with the pressure of the confining neutral gas, i.e., 

\begin{equation}
n_{\rm H}(r=R) \simeq \frac{p_{\rm edge}}{kT_{\rm PDR}}.
\end{equation}

\noindent For the PDR temperature we assume $T_{\rm PDR}=500$~K\footnote{Assuming $T_{\rm PDR}=300$~K instead of 500~K  produces very similar results. Assuming $T_{\rm PDR}=1000$~K results in \cii/FIR ratios about $50\%$ lower for $\Sigma_{\rm FIR}\lesssim10^{10}$~$L_{\odot}$~kpc$^{-2}$.}, consistent with observed temperatures in Galactic PDRs \citep{rhc_hollenbach99,rhc_sheffer11}.

Now that we have expressions for \goob\ and \nh\ as a function of $R$, we consider a family of \hii\ region/molecular cloud complexes with sizes that range from $R=2$~pc to 100~pc, and for each size we use PDR Toolbox 
\citep{rhc_kaufman06,rhc_pound08} to compute the intensity of the \cii, \oi, and \oii\ lines (in units of erg~cm$^{-2}$~s$^{-1}$~sr$^{-1}$). Following \cite{rhc_kaufman99}, we also calculate the FIR dust continuum intensity in the optically-thin limit as $I_{\rm FIR}=2\times1.3\times10^{-4}$\goob\ [erg~cm$^{-2}$~s$^{-1}$~sr$^{-1}$] (recall that \gof\ corresponds to $1.6\times10^{-3}$~erg~cm$^{-2}$~s$^{-1}$ and $1.3\times10^{-4}=1.6\times10^{-3}/4\pi$), although we note that the optically-thin limit assumption is not accurate for Arp 220-like environments characterized by extinction of the FIR lines due to dust and high column densities of OH and H$_{2}$O \citep{rhc_gonzalez-alfonso04,rhc_gonzalez-alfonso12,rhc_gonzalez-alfonso15}. Finally, we calculate the FIR surface brightness as $\Sigma_{\rm FIR} \simeq 2.5\times10^9\times4\pi I_{\rm FIR}\times\Phi_{\rm A}~[L_{\odot}$~kpc$^{-2}$], where $\Phi_{\rm A}$ is the beam area filling factor of the PDR regions. When $\Phi_{\rm A}<1$, the area of \cii\ emitting sources does not fill the beam, while $\Phi_{\rm A}>1$ means that the PDR surface area intercepted by the beam exceed the beam projected area.

\begin{figure*}
\begin{center}
\includegraphics[scale=0.23]{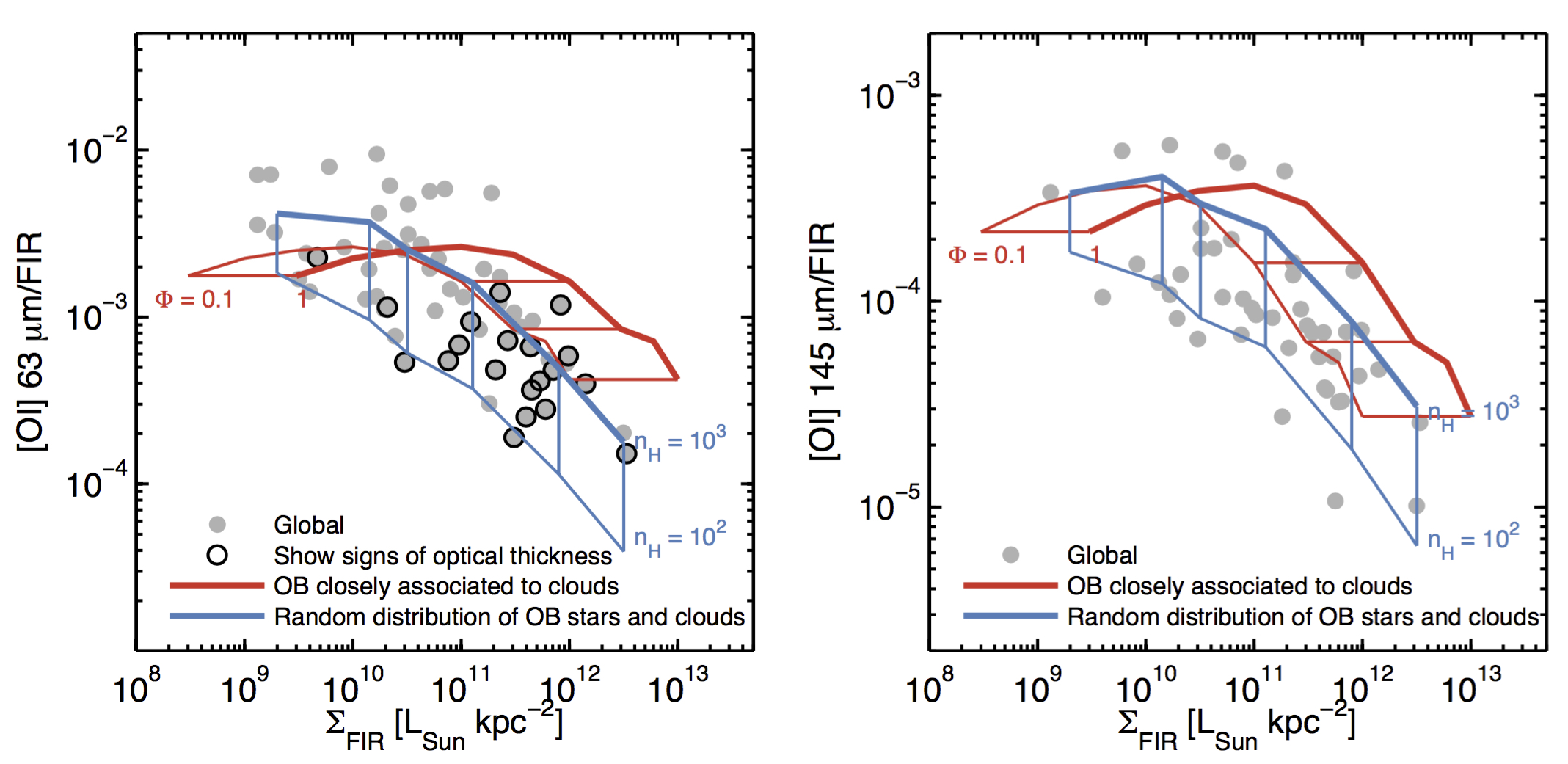}
\caption{\oi-to-FIR (left) and \oii-to-FIR ratio (right) as a function of \sfir. Global  measurements of nearby galaxies are shown as grey circles. We overplot the model results for both scenarios: (1) dense PDRs varying the beam area filling factor $\Phi$ from 0.1 to 1 (red lines), and (2) random distribution of OB stars and clouds varying the neutral gas density \nh\ from 10$^2$~cm$^{-3}$ to 10$^3$~cm$^{-3}$ (blue lines). Galaxies that show some degree of optical thickness in their \oi\ line emission (identified as sources with \ois~63/145~$\mu$m line ratios $<10$, and assuming that the \oii\ line emission is optically thin) are marked with a black border.
\label{Fig:oi}}
\end{center}
\end{figure*}

In this scenario we consider that OB stars are randomly placed with respect to the clouds. In this case the FUV flux incident on clouds is dominated by the average interstellar radiation field (ISRF). Following \cite{rhc_wolfire90}, we can express \gof\ in this particular scenario as:

\begin{equation}
\begin{split}
\langle G_{0} \rangle^{\rm ISRF} \simeq 3\times10^4\Bigg(\frac{L_{\rm FIR}}{10^{10}~L_{\odot}}\Bigg)\times\Bigg(\frac{\lambda}{100~{\rm pc}}\Bigg) \\
\times\Bigg(\frac{100~{\rm pc}}{S}\Bigg)^3[1-e^{-S/\lambda}],
\end{split}
\end{equation}

\noindent where $S$ is the radius of the IR emitting region and $\lambda$ is the mean cloud separation along the line of sight.

For the neutral gas volume density of the surface of the clouds illuminated by \goisrf, we  assume values of $10^2$ and 10$^3$~cm$^{-3}$, which represent the range of observed densities in our Galaxy and extragalactic sources \citep[e.g.,][]{rhc_stacey85,rhc_wolfire10,rhc_gracia-carpio11,rhc_pineda13}. Then, assuming an IR source with a luminosity of $L_{\rm IR}=5\times10^{10}$~$L_{\odot}$, and varying the radius of the IR emitting source from $S=0.1$~kpc to 2~kpc (which produces FIR surface brightnesses in the  $\Sigma_{\rm FIR}\sim10^9-10^{12}$~$L_{\odot}$~kpc$^{-2}$ range), we calculate \goisrf\ assuming that the mean separation between clouds is twice the mean separation between stars (i.e., $\lambda=2d_{*}\approx2\times[(L_{\rm FIR}/L_{*})/\frac{4}{3}\pi S^3]^{-1/3}$; for the luminosity of the star cluster we assume $L_{*}=5\times10^5$~$L_{\odot}$, the typical value for OB associations in clouds such as Orion or M17). Finally, for each size $S$ and its corresponding \goisrf\ value, we use PDR Toolbox to calculate the intensity of the \cii\ and \ois\ lines for assumed characteristic densities of $n_{\rm H}=10^2$ and 10$^3$~cm$^{-3}$ \citep[e.g.,][]{rhc_malhotra01,rhc_parkin13,rhc_contursi17}.

As a final note on the model, calculations in the PDR Toolbox code assume a grain photoelectric heating rate as derived by \cite{rhc_bakes94}.  This rate includes a size distribution of particles extending from large grains ($\sim0.25~\mu$m) to small ($\sim5$~\AA) PAHs and explicitly accounts for the microphysics of small particles. In the numerical calculations of \cite{rhc_bakes94}, the grain photo-electric heating efficiency, $\epsilon_{ph}$, is a function of the charging parameter, $\gamma=G_{0}T^{1/2}/n_{\rm e}$. We calculate $\gamma$ using $T$ and $G_{0}$ given by our assumption of model scenario. For the electron density we use the analytic expression presented in \cite{rhc_roellig06}\footnote{The analytic expression in \cite{rhc_roellig06} for the electron density is: $n_{\rm e} \approx 0.84\times10^{-4} \times nZ(1+(1+14.4T^{0.75}/(nZ^2))^{1/2})~{\rm cm}^{-3}$. The value of $T$ and $n$ are given by our model, and we assume gas with solar metallicity.}. 

\subsubsection{Toy Model results and comparison to observations} 

When comparing the line-to-dust continuum ratios observed in galaxies to those calculated by our model, we need to consider that the latter does not include contributions to the line emission by the ionized phase. This is of particular importance for the \cii\ line that can arise from both the ionized and the neutral gas. One observational tool to constrain the fraction of the \cii\ emission emitted by the neutral gas, $f_{\rm [CII]}^{\rm neutral}$, is to compare the \cii\ to 
one of the \nit\ lines. As discussed in Paper~I \citep{rhc_rhc18a}, based on the \cii/\nii\ ratio we find that $f_{\rm [CII]}^{\rm neutral}$ increases from $\sim60\%$ to 90\% in the \sratio$\sim0.1-2$ range.

The neutral gas components contributing to the \cii\ emission are PDRs, diffuse atomic gas, and ``CO-faint'' molecular gas (i.e., molecular H$_{2}$ gas that resides in parts of clouds where CO has been dissociated). For the close association of stars and clouds we assume that the \cii\ emission is dominated by the PDRs and ``CO-faint'' molecular gas as indicated by the analysis of normal and starburst galaxies. \citep[e.g.,][]{rhc_kaufman06,rhc_stacey10,rhc_croxall12,rhc_pineda14}. This is also the case for the \ois\ lines, as standard models of starlight heating indicate that diffuse gas is too cool to emit \ois. 

Figure~\ref{Fig:cii} shows the comparison between our model outputs and the observed \cii-to-FIR ratio as a function of \sfir. We have scaled the model outputs by the inverse of $f_{\rm [CII]}^{\rm neutral}$ in order to account for the contribution from the ionized gas to the \cii\ emission.  We calculate $f_{\rm [CII]}^{\rm neutral}$ using a parameterization of this fraction as a function of \sfir\ \citep{rhc_croxall17,rhc_diaz-santos17} where $f_{\rm [CII]}^{\rm neutral}$ increases from $\sim70\%$ to $\sim90\%$ in the $\Sigma_{\rm FIR}\sim10^{9}-10^{12}$~$L_{\odot}$~kpc$^{-2}$ range. 

The red curve shows the model result for the close-association model assuming beam filling factors of $\Phi_{A}=0.1,0.3$ and 1. Starburst galaxies and LIRGs typically have beam filling factors in the $\Phi_{A}\simeq10^{-2}-1$ range; starburst galaxies with \sfir\,$\gtrsim10^{11}$~$L_{\odot}$~kpc$^{-2}$ tend to have beam filling factors approaching unity \citep[e.g., ][]{rhc_diaz-santos17}. In the dense PDR scenario, as \hii\ regions become more compact \gof\ increases as $R^{-2}$; \nh\ also increases but at a slightly lower rate. If the cloud density \nh\ is lower than the critical density ($n_{\rm crit}$) of the \cii\ line --which in our model occurs when  $\Sigma_{\rm FIR}\lesssim3\times10^{10}$~$L_{\odot}$~kpc$^{-2}$-- then the intensity of \cii\ emission is nearly independent of \gof\ and only increases proportional to $n_{\rm H}$ \citep{rhc_kaufman99}. This implies that as \hii\ regions become more compact, the FIR dust continuum intensity ($\propto G_{0}$) increases at a faster rate than  the \cii\ intensity ($\propto n_{\rm H}$), causing the \cii/FIR ratio to decrease as a function of increasing \sfir. Once the gas density of the neutral gas cloud confining the \hii\ region reaches the critical density of the \cii\ transition at around $\Sigma_{\rm FIR}\sim3\times10^{10}$~$L_{\odot}$~kpc$^{-2}$ (for $\Phi=1$), the intensity of \cii\ emission becomes independent of both \gof\ and \nh, resulting in the decrease of the \cii/FIR ratio with \sfir\ at an even faster rate. In terms of the PDR mass --which is proportional to $N_{\rm C^+}R^2$--, as the size of the \hii\ region decreases $N_{\rm C^+}$ remains constant but \gof, and consequently \sfir, increases resulting in lower PDR masses available to produce \cii\ emission \citep[see also][]{rhc_gonzalez-alfonso08,rhc_gonzalez-alfonso15}.

One additional factor in the dense PDR scenario that can contribute to the decrease of the \cii-to-FIR ratio is the competition for the available UV photons between dust, ionized gas and neutral hydrogen. The expectation is that as the ionization parameter $U$ increases, the fraction of UV photons that ionize and excite the gas is reduced by dust absrorption, which results in enhanced FIR continuum emission compared to the intensity of the FIR lines \citep[e.g.,][]{rhc_voit92,rhc_luhman03,rhc_abel09,rhc_gracia-carpio11,rhc_fischer14}. In the \citealt{rhc_draine11} model of dusty \hii\ regions the compression of the ionized gas into a shell limits the ionization parameter at the half-ionization radius to $U\approx0.01$ \citep[see also][]{rhc_dopita02, rhc_stern14}. This value corresponds to the threshold where we expect dust to start absorbing an important fraction of the incident ionizing radiation \citep[e.g.,][]{rhc_gracia-carpio11}, thus reducing the line-to-FIR ratio. This, in conjunction with the density effect described above, may also contribute to the decrease in the \cii-to-FIR ratio as a function of \sfir\ in the dense PDR scenario.

Finally, in the dense PDR scenario the drop of the [CII]/FIR ratio as a function of \sfir\ is not a result of the decrease in the photoelectric heating efficiency. In the $\Sigma_{\rm FIR}\sim10^9-10^{12}$~$L_{\odot}$~kpc$^{-2}$ range, the charging parameter increases from $\gamma\sim1\times10^{4}$ to $\sim8\times10^{4}$~K$^{1/2}$~cm$^{-3}$, which implies only a modest reduction in the [CII]/FIR ratio of a factor of $\sim4$. These values of $\gamma$ are consistent with those found in the HII/PDR complexes such as NGC~7023, Mon~R2, and Ced~201 \citep{rhc_okada13}.

In the second scenario we consider an average interstellar radiation field \goisrf\ illuminating neutral clouds with densities that vary from $n_{\rm H}=10^2$~cm$^{-3}$ (thin blue line) to $10^3$~cm$^{-3}$ (thick blue line). Around $\Sigma_{\rm FIR}\approx10^{9}$~$L_{\odot}$~kpc$^{-2}$, the average interstellar radiation field is $\langle G_{0} \rangle^{\rm ISRF}\approx6$ and the  efficiency of gas heated by photoelectric heating is high (${\rm [CII]/FIR\sim3\%}$), but still in the range of values observed in systems such as low-metallicity galaxies \citep[e.g.,][]{rhc_cigan16}. Now, as we make galaxies more compact, the average interstellar radiation field starts to increase, and at around $\Sigma_{\rm FIR}\approx10^{10}$~$L_{\odot}$~kpc$^{-2}$ reaches $\langle G_{0} \rangle^{\rm ISRF}\approx10^{2}$. From this point on, and if we keep $n_{\rm H}$ fixed (at values lower than $n_{\rm crit}$), the intensity of \cii\ becomes  nearly independent of \gof\ as this transition saturates at gas temperatures above 92~K (i.e., increases in $T$ do not appreciably change the intensity) and proportional to $n_{\rm H}$. At $\Sigma_{\rm FIR}\approx3\times10^{11}$~$L_{\odot}$~kpc$^{-2}$ the mean separation between clouds according to the model is $\lambda\approx10$~pc, and the \cii-to-FIR ratio is $\approx10^{-3}$. These numbers are consistent with those observed in the starbursting region of M~82 where $\lambda\approx2-7$~pc \citep{rhc_lord96,rhc_nfs01} and ${\rm [CII]/FIR}\approx1-3\times10^{-3}$ \citep[This work and][]{rhc_contursi13}.
Regarding the efficiency of the photoelectric heating, as $\Sigma_{\rm FIR}$ increases from $\Sigma_{\rm FIR}\sim10^9$ to $10^{12}$~$L_{\odot}$~kpc$^{-2}$, the charging parameter increases from $\gamma\sim1\times10^{3}$ to $\sim8\times10^{6}$~K$^{1/2}$~cm$^{-3}$~\footnote{These minimum and maximum values of $\gamma$ are similar to the values found in the Horsehead nebula and the NW component of NGC~7023, respectively \citep{rhc_okada13}}. This nearly four orders of magnitude increment in the charging parameter implies a reduction in the photoelectric heating efficiency, and hence the [CII]/FIR ratio, of a factor of $\sim500$ due to the charging of the dust grains. The net result is that for a fixed neutral cloud density the \cii-to-FIR ratio decreases as $G_{0}/n_{\rm H}$, and consequently \sfir, increases. 

\begin{figure*}
\begin{center}
\includegraphics[scale=0.23]{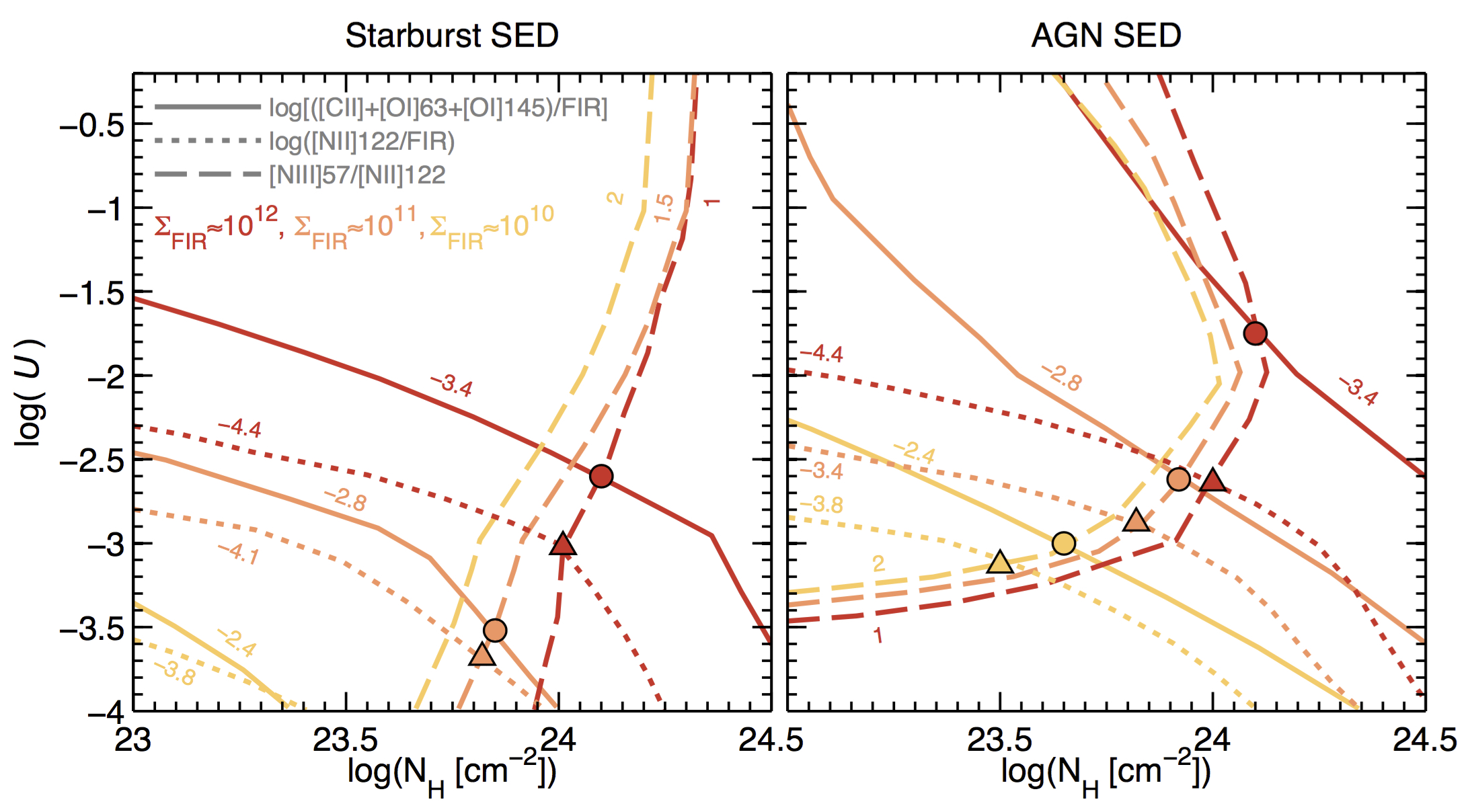}
\caption{{\it Cloudy} predictions for the (\cii+\oi+\oii)/FIR (solid contours), \nii/FIR (dotted contours), and \niii/\nii\ (dashed contours) ratios as a function of hydrogen column density (\NH) and ionization parameter ($U$) assuming starburst (left) or AGN (right) central illumination SEDs \citep{rhc_fischer14}. The contour values are chosen to represent the mean values measured in galaxies with $\Sigma_{\rm FIR}\approx10^{10}$~$L_{\odot}$~kpc$^{-2}$ (yellow), $\Sigma_{\rm FIR}\approx10^{11}$~$L_{\odot}$~kpc$^{-2}$ (orange), and $\Sigma_{\rm FIR}\approx10^{12}$~$L_{\odot}$~kpc$^{-2}$ (red). For each galaxy group we mark with a circle the intersection between the (\cii+\oi+\oii)/FIR and \niii/\nii\ curves, and with a triangle the intersection between the \nii/FIR and \niii/\nii\ curves. The overlap regions of the model results shows that the observed line deficit of the PDR lines can be explained as a result of the increase in the hydrogen column density and the ionization parameter. \label{Fig:fm14}}
\end{center}
\end{figure*}

The majority of the \cii-to-FIR ratios observed in our galaxies lie between the dense PDR and average ISRF model curves. This is expected as, in reality, the ISM structure is a combination of these two scenarios. For example, \cite{rhc_parravano03} find that averaged over the lifetime of a massive cluster, about $\sim20\%$ of the FUV photons produced by massive stars end up illuminating nearby dense molecular clouds, while $\sim80\%$ reaches the diffuse ISM. In the Orion molecular complex, \cite{rhc_goicoechea15} find that $\sim85\%$ of the total \cii\ luminosity arises from a dense PDR component (with typical \cii/FIR ratios in the $\sim10^{-4}-10^{-3}$ range, $L_{\rm FIR}/M_{\rm H_2}\gtrsim80$~$L_{\odot}~M_{\odot}^{-1}$, $G_{0}\gtrsim10^4$ and $n_{\rm H}\gtrsim10^5$~cm$^{-3}$) and an extended cloud component (with \cii/FIR ratios in the $\sim10^{-3}-10^{-2}$ range).

The threshold value of $\Sigma_{\rm FIR}\sim3\times10^{10}$~$L_{\odot}$~kpc$^{-2}$ at which our {\it toy model} predicts a decrease in the \cii/FIR ratio is remarkably similar to the \sfir\ value that separates galaxies with normal modes of star formation from compact starbursts  \citep{rhc_elbaz11}, and PDR regimes having constant or increasing $G_{0}/n_{\rm H}$ ratios as a function of \sfir \citep{rhc_diaz-santos17}. In addition, SHINING galaxies with $\Sigma_{\rm FIR}\gtrsim3\times10^{10}$~$L_{\odot}$~kpc$^{-2}$ tend to have \firmol$\gtrsim50$~$L_{\odot}~M_{\odot}^{-1}$, which is characteristic of galaxies with a faster and/or more efficient mode of star formation \citep[e.g.,][]{rhc_genzel10,rhc_daddi10,rhc_gracia-carpio11}. This suggests that the mean properties of PDRs and \hii\ regions in these two groups of galaxies are different. Galaxies with $\Sigma_{\rm FIR}\gtrsim3\times10^{10}$~$L_{\odot}$~kpc$^{-2}$ have more compact, dense PDR/\hii\ region complexes powered by a higher fraction of young, massive stars that produce harsh radiation fields in their vicinity, and a higher average radiation field in the disk. All these combined produce global \cii/FIR ratios that are lower compared to those in galaxies with normal and more extended star formation activity.

As a cautionary note, it is important to mention that additional physical effects not considered in our models can also contribute to the \cii\ deficit, including \cii\ self-absorption and high dust opacity in very obscured, dense starbursts \citep[e.g, Arp~220; ][]{rhc_gonzalez-alfonso04,rhc_rangwala11,rhc_scoville17}, or the impact of a very powerful AGN on the ionization state of the gas \citep[e.g,][]{rhc_langer15}. Regarding other models and simulations that address the problem of the \cii\ deficit we refer to \cite{rhc_abel09}, \cite{rhc_fischer14}, \cite{rhc_gonzalez-alfonso08,rhc_gonzalez-alfonso15}, \cite{rhc_munoz15}, \cite{rhc_langer15}, \cite{rhc_narayanan16}, and \cite{rhc_diaz-santos17}. In particular, \cite{rhc_gonzalez-alfonso15} use a composite model that simultaneously describes the OH absorption, the high \firmol\ ratios, and the \cii\ deficit observed in local (U)LIRGs. The model consists of a warm, optically thick component that is responsible for the molecular-absorption-dominated spectra in (U)LIRGs and that emits inefficiently in \cii\ emission, and a colder, optically thin component with a fix density of $n_{\rm H}=10^3$~cm$^{-3}$ that produces the bulk of the \cii\ emission. In this model the \cii\ deficit --that is correlated with the equivalent width of the absorbing OH 65~$\mu$m line and \firmol-- is a consequence of the limited reservoir of C$^{+}$ ions that at high luminosity-to-gas mass ratios limits the \cii\ luminosity per unit of luminous power in the FIR continuum \citep[see for example Equation~5 in][]{rhc_gonzalez-alfonso08}.

Figure~\ref{Fig:oi} shows the comparison between our models and the observed \ois-to-FIR ratios as a function of \sfir. In the case of \oi\ (left panel), both models overlap and agree reasonably well with the observations until reaching $\Sigma_{\rm FIR}\approx3\times10^{10}$~$L_{\odot}$~kpc$^{-2}$. Above this surface brightness the model predictions differ: for clouds irradiated by an average ISRF, $\langle G_{0} \rangle^{\rm ISRF}$ reaches $\sim400$, which results in the high-temperature saturation of the \oi\ line and the decline of the \oi-to-FIR ratio with \sfir. On the other hand, in the dense PDR formulation we do not predict a decline in the \oi-to-FIR ratio until we reach IR surface brightnesses of the order of $\Sigma_{\rm FIR}\approx5\times10^{11}$~$L_{\odot}$~kpc$^{-2}$. This threshold \sfir\ value is significantly higher than in the \cii\ case due to the fact that the critical density of the \oi\ line ($n_{\rm crit}=4.7\times10^5$~cm$^{-3}$) is a factor $\times150$ higher than the critical density of the \cii\ line.

At first sight, it seems that the decline in the observed \oi-to-FIR ratio is better described by the scenario of clouds irradiated by an average ISRF. However, when interpreting \oi\ observations it is important to consider that this line can become optically thick \citep[e.g.,][]{rhc_poglitsch96,rhc_malhotra01,rhc_farrah13,rhc_rosenberg15} and suppressed through
self-absorption and dust extinction \citep[e.g.,][]{rhc_luhman03,rhc_gonzalez-alfonso04,rhc_gonzalez-alfonso08,rhc_gonzalez-alfonso12,rhc_vasta10}. In general, galaxies with \ois~63/145~$\mu$m ratios lower than $\sim10$ (assuming that \oii\ is optically thin) can be considered to exhibit some degree of optical thickness in the \oi\ emission. On Figure~\ref{Fig:oi} we mark galaxies in this category using black thick circles. These systems correspond to the brightest galaxies in our sample, and we can not discard that their un-absorbed \oi-to-FIR ratios are higher and in better agreement with the dense PDR model results. 

Finally, the right panel of Figure~\ref{Fig:oi} shows the results for the \oii\ line. Overall there is a good agreement between the model results and the observed \oii-to-FIR ratios. The characteristics of the model outputs are also similar to that of the \oi\ line, except that in the dense PDR scenario the \oii-to-FIR ratio stars to drop at a lower \sfir\ value. The reason is that the critical density of the \oii\ transition is a factor of $\sim5$ lower than the critical density of the \oi\ line. 

\subsection{Cloudy modeling of the observed infrared line ratio trends \label{cloudy}}

In addition to the analysis of the line-to-continuum trends based on our {\it toy model}, in this section we continue with the interpretation of the observed line ratios using the framework built by \cite{rhc_abel09} and \cite{rhc_fischer14} based on the {\it Cloudy} spectral synthesis code \citep{rhc_ferland99,rhc_ferland13}. 

The models \citep[described in detail in][]{rhc_fischer14} consider a spherical one-dimensional geometry where the central source of heating is dominated by a young starburst or an AGN. {\it Cloudy} is used to compute the thermal and chemical structure of the gas cloud from the illuminated surface of hot, ionized hydrogen into regions with high hydrogen column density ($N_{\rm H}$) and optical extinction ($A_{\rm V}$) where atoms have combined into molecules. The ionized gas density of the illuminated face of the cloud was set by \cite{rhc_fischer14} to $n_{\rm H^+}=30, 300$, or 3000~cm$^{-3}$, and calculations of the line intensities and dust continuum are measured as a function of the ionization parameter ($U$) up to hydrogen column densities of $N_{\rm H}=10^{25}$~cm$^{-2}$ ($A_{\rm V}=4000$~mag). The advantage of using {\it Cloudy} is that the code incorporates in one model several processes relevant to the modeling of the ISM conditions, including photoionization and photodissociation, cosmic-ray ionization and heating, photoelectric heating of gas as a function of dust grain properties, and the effect of thermal, radiation, and magnetic pressure. On the other hand, one limitation of the simple geometry assumed by the model is that emission from extra-nuclear regions is not included, which can be relevant for galaxies where star formation is distributed throughout the disk, rather than at a single central position.

To facilitate the comparison between the observed line ratios and the model predictions, we divide our galaxies into three groups based on their FIR surface brightness ($\Sigma_{\rm FIR}\geq10^{12}$~$L_{\odot}$~kpc$^{-2}$,  $10^{11}\leq\Sigma_{\rm FIR}<10^{12}$~$L_{\odot}$~kpc$^{-2}$, and  $\Sigma_{\rm FIR}<10^{11}$~$L_{\odot}$~kpc$^{-2}$). These three categories represent groups of galaxies with line deficits that range from strong to moderate. For each group we measure the mean (\cii+\oi+\oii)/FIR, \nii/FIR, and \niii/\nii\ ratios. These are proxies for the cooling budget and the gas heating efficiency, the ionizing photon flux, and the hardness of the UV radiation field, respectively. We find that as the FIR surface brightness of galaxies increases from $\Sigma_{\rm FIR}\sim10^{10}$ to $\sim10^{12}$~$L_{\odot}$~kpc$^{-2}$, the mean (\cii+\oi+\oii)/FIR, \nii/FIR, and \niii/\nii\ ratios drop by factors of $\sim10$, $\sim4$, and $\sim2$, respectively. 


Figure~\ref{Fig:fm14} shows the model contours representing the mean ratios for the three galaxy categories as a function of $U$ and $N_{\rm H}$. We assume an electron density for the illuminated face of H$^{+}$ gas of $n_{\rm H^+}=300$~cm$^{-3}$, in agreement with observed electron densities in central kilo-parsec size regions with $\Sigma_{\rm FIR}\sim10^{10-11}$~$L_{\odot}$~kpc$^{-2}$ \citep[][]{rhc_rhc16}. The left and right panels show the model predictions when assuming a starburst or AGN central source, respectively. In both cases the observed drop in the (\cii+\oi+\oii)/FIR and \nii/FIR ratios can be explained by increasing both \NH\ and $U$. The \niii/\nii\ ratio, on the other hand, is not very sensitive to changes in the ionization parameter, and its orthogonality with respect to the other line-to-continuum ratios can be used to constrain the characteristic values of \NH\ and $U$ for each galaxy group. In the figure we mark with circles (squares) the position in the $U$, \NH\ plane where the (\cii+\oi+\oii)/FIR (\nii/FIR) and the \niii/\nii\ contours intersect. 

In the starburst dominated scenario, the intersection of the contours for the $\Sigma_{\rm FIR}\sim10^{10}$~$L_{\odot}$~kpc$^{-2}$ galaxies falls outside the probed parameter space, but the trends seems to indicate that  the contours will overlap around \NH$\sim10^{23.5}$~cm$^{-2}$ and $U\sim10^{-4.2}$. For galaxies with $\Sigma_{\rm FIR}\sim10^{11}$~$L_{\odot}$~kpc$^{-2}$ we find characteristic values of \NH$\sim10^{23.8}$~cm$^{-2}$ and $U\sim10^{-3.5}$, and for the $\Sigma_{\rm FIR}\sim10^{12}$~$L_{\odot}$~kpc$^{-2}$ group we find \NH$\sim10^{24.1}$~cm$^{-2}$ and $U\sim10^{-2.7}$. For an AGN central SED, we observe that the decrease in the relative intensity of the PDR and \nii\ lines of galaxies in the $\Sigma_{\rm FIR}\sim10^{10-12}$~$L_{\odot}$~kpc$^{-2}$ range can be explained by increasing \NH\ from $\sim10^{23.5}$ to $\sim10^{24}$~cm$^{-2}$ and $U$ from $\sim10^{-3}$ to $\sim10^{-2}$. 

In conclusion, we find that regardless of the choice of central power source, the models can reproduce the PDR and \nii\ line deficits observed in galaxies in the $\Sigma_{\rm FIR}\sim10^{10-12}$~$L_{\odot}$~kpc$^{-2}$ range by increasing both \NH\ and $U$ by factors of $\sim4$ and $\sim10$, respectively. This is consistent with the results found by \cite{rhc_gracia-carpio11}. The high-column densities (\NH$\sim10^{24}$~cm$^{-2}$) predicted for the $\Sigma_{\rm FIR}\sim10^{12}$~$L_{\odot}$~kpc$^{-2}$ galaxies are compatible with the low mid-IR line to FIR continuum ratios \citep{rhc_farrah13} and high OH and H$_{2}$O column densities \citep[e.g.,][]{rhc_gonzalez-alfonso12,rhc_gonzalez-alfonso15} observed in these systems. In addition, the high-$U$ values are required to maintain the observed warm FIR colors \citep[see also][]{rhc_fischer14}. This {\it Cloudy} model interpretation of our results is also consistent with our {\it toy model} calculations. In both cases we find that strong-line deficits are associated with gas clouds with $U\sim10^{-2}$. At this high $U$ value the absorption of UV photons by dust in the ionized region is significant, plus the density of the gas can exceed the critical density of the \cii\ line \citep[e.g., our {\it toy model} results or Figure~1 in][]{rhc_abel09}.

\subsection{The influence of AGN activity on the \cii/FIR ratio}\label{section:AGN}

\begin{figure*}
\begin{center}
\includegraphics[scale=0.21]{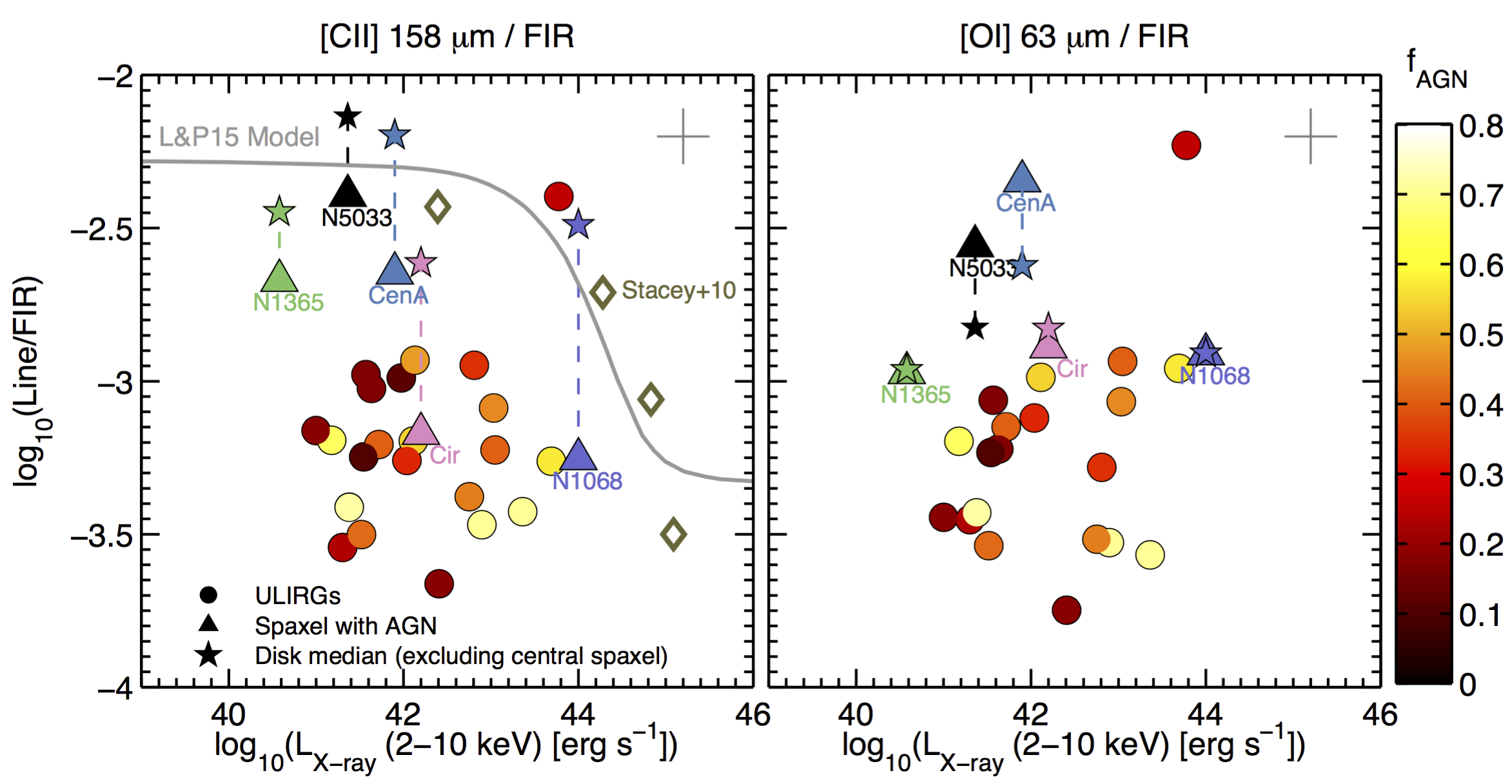}
\caption{{\it (Left)} \cii/FIR ratio as a function of intrinsic (absorption-corrected) X-ray luminosity ($2-10$~keV) of ULIRGs (color circles), the central spaxel (triangles) and the median of the disk excluding the central spaxel (stars) of 5 resolved SHINING AGN galaxies, and galaxies in the \cite{rhc_stacey10} sample (empty dark green diamonds). For the ULIRGS, the color scale indicates the fractional contribution of nuclear activity to the bolometric luminosity ($f_{\rm AGN}$) as calculated by \cite{rhc_veilleux09}. The gray solid line represents the model predictions by \cite{rhc_langer15} assuming a PDR filling factor of 0.002. {\it (Right)} Similar to the left panel, but this time the ordinate indicates the \oi/FIR ratio. A typical errorbar is plotted in the upper-right corner.}\label{cii_agn}
\end{center}
\end{figure*}

X-ray dominated regions (XDRs) produced by AGNs heat the surrounding gas and dust, and this can have a significant impact on the (local) \cii/FIR ratio \citep[e.g.,][]{rhc_maloney96,rhc_meijerink05}. In particular, harsh radiation fields present in XDRs can decrease the cooling power of \cii\ by: 
(1) destroying small dust grains and PAHs, thus reducing the photoelectric heating efficiency of the gas \citep{rhc_voit92}; (2) converting a fraction of the C$^{+}$ ions to higher ionization states \citep{rhc_langer15}; and (3) heating the dust and gas to temperatures high enough for the \oi\ line to become the dominant cooling channel \citep[e.g.,][]{rhc_abel09}.

To explore the connection between the AGN power and its influence on the dust continuum and PDR line emission, in Figure~\ref{cii_agn} we plot the \cii/FIR (left) and \oi/FIR (right) ratio as a function of the X-ray luminosity ($L_{\rm X-ray}$) for ULIRGS, $z\sim1$ galaxies from \cite{rhc_stacey10} (3 AGN dominated, 1 mixed, and 2 starburst dominated systems), and resolved regions in Seyfert SHINING galaxies. The absorption-corrected  X-ray luminosities in the $2-10$~keV range were retrieved from \cite{rhc_iyomoto97,rhc_cappi06,rhc_teng10,rhc_derosa12}. The gray solid line shows the model predictions from \cite{rhc_langer15} for the \cii/FIR ratio as a function of X-ray luminosity\footnote{The X-ray luminosities computed in the \cite{rhc_langer15} model are integrated for energies larger than 1~keV, while the X-ray luminosities of the galaxies included in Figure~\ref{cii_agn} cover the $2-10$~keV range. As described by \cite{rhc_langer15}, if we assume an X-ray photon index of $\Gamma=1.9$ this implies that data points should shift $\sim0.6$~dex to the right.}. In this model, the \cii\ emission arises from dense warm ionized gas and PDRs, and the main effect of the presence of a strong source of X-ray flux is to convert a fraction of the C$^{+}$ ions to higher ionization states, thus reducing the \cii\ luminosity. They predict this effect should reduce the \cii/FIR ratio by a factor of a few at $L_{\rm X-ray}\approx10^{44}$~erg~s$^{-1}$, and by an order of magnitude at $L_{\rm X-ray}\gtrsim10^{45}$~erg~s$^{-1}$.

To study the local effect of the AGN on the \cii/FIR ratio, we use the SHINING Seyfert galaxies that are spatially resolved to compare the \cii/FIR ratio measured in the central spaxel versus the median of the disk (excluding the central spaxel). For the four galaxies with $L_{\rm X-ray}\lesssim10^{42}$~erg~s$^{-1}$ (NGC~1365, NGC~5033, Centaurus~A, and Circinus), we observe a central \cii/FIR ratio a factor of $\sim2-3$ lower compared to the disk's median. Interestingly, the central regions in these systems do not show an \oi/FIR ratio deficit relative to the disk, in fact, NGC~5033 and Centaurus~A have central \oi/FIR ratios a factor of $\sim2$ higher than the median of their disks. This suggests that the suppression of the central \cii/FIR ratio in these Seyfert galaxies is a consequence of the dominant contribution of the \oi\ line to the cooling of the neutral gas.

The remaining SHINING Seyfert galaxy that is spatially resolved is the archetypical system NGC~1068. The \cii/FIR ratio measured in the central spaxel (spatial scale $\sim500$~pc) is $6\times10^{-4}$, a value that is typical of those observed in ULIRGs, and a factor of $\sim6$ lower than the median ratio measured in its disk. Similar to the other Seyfert galaxies in our sample, we do not observe central suppression of the \oi/FIR ratio relative to the disk. As a consequence, the cooling of the neutral gas in the central spaxel of NGC~1068 is dominated by the \oi\ line (the \oi/\cii\ line luminosity ratio is 2.3). The strong central X-ray source in NGC~1068 place this galaxy in the regime where the \cite{rhc_langer15} model predicts that changes in the overall state ionization of the gas can contribute to the \cii/FIR suppression. Additional evidence that supports this scenario is that the ratio between the \oiii\ line (originated in highly ionized gas) and the \cii\ line in the central spaxel is \oiii/\cii~$=1.3$, one of the highest values in the entire sample of resolved and un-resolved SHINING galaxies \citep[see Figure~11 in Paper~I; ][]{rhc_rhc18a}.

\begin{figure}
\begin{center}
\includegraphics[scale=0.12]{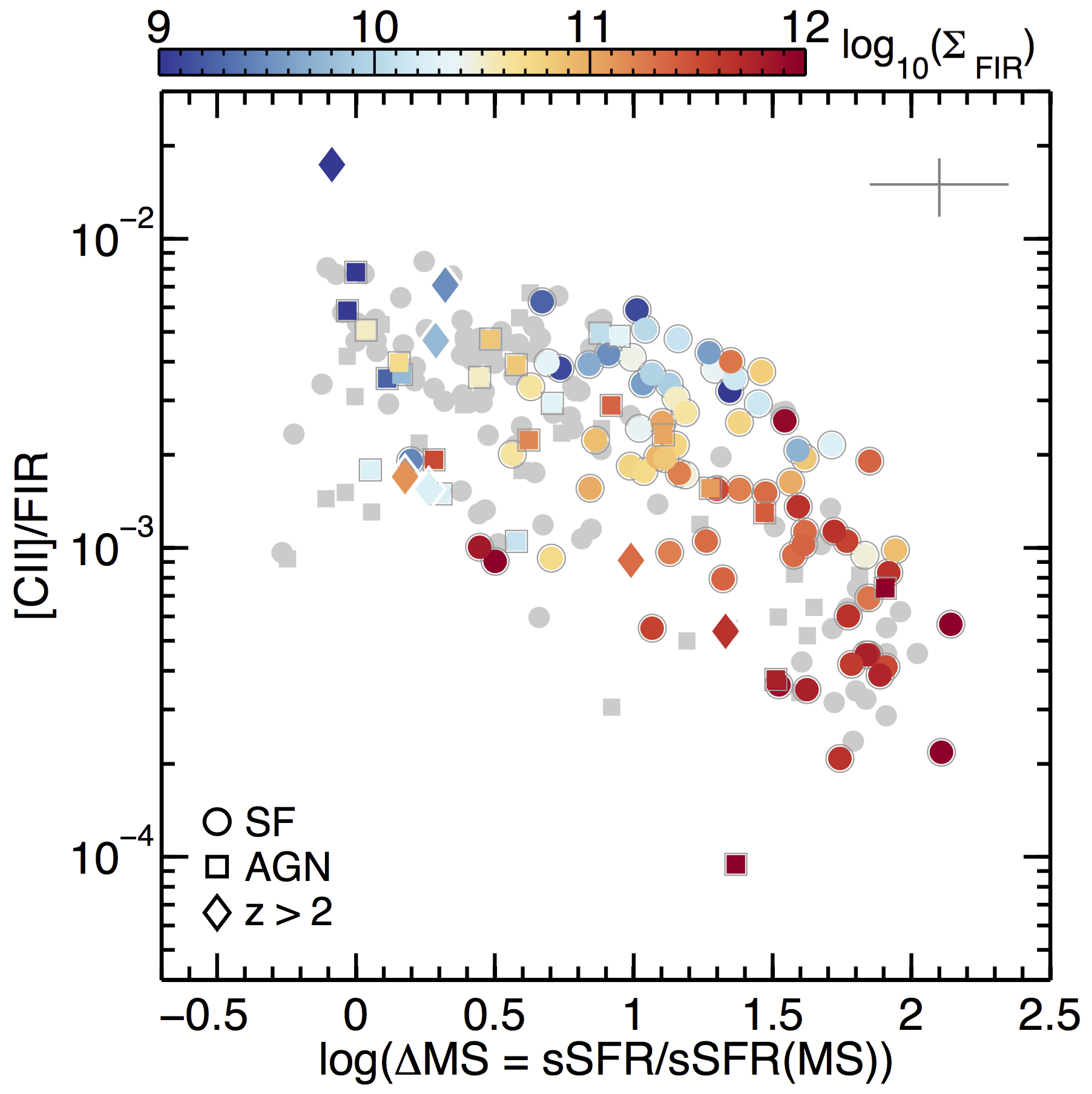}
\caption{\cii/FIR ratio as a function of specific star-formation rate normalized to the mid-line of main sequence \citep{rhc_speagle14} as a function of redshift ($\Delta{\rm MS}$). The color scale indicates the FIR surface brightness of the galaxies for which we have size measurements available. Star-forming galaxies are shown as circles, AGNs as squares, and high-$z$ ($z>2$) systems  galaxies as diamonds. A typical errorbar is plotted in the upper-right corner.}\label{cii_MS}
\end{center}
\end{figure}

\begin{figure*}
\begin{center}
\includegraphics[scale=0.26]{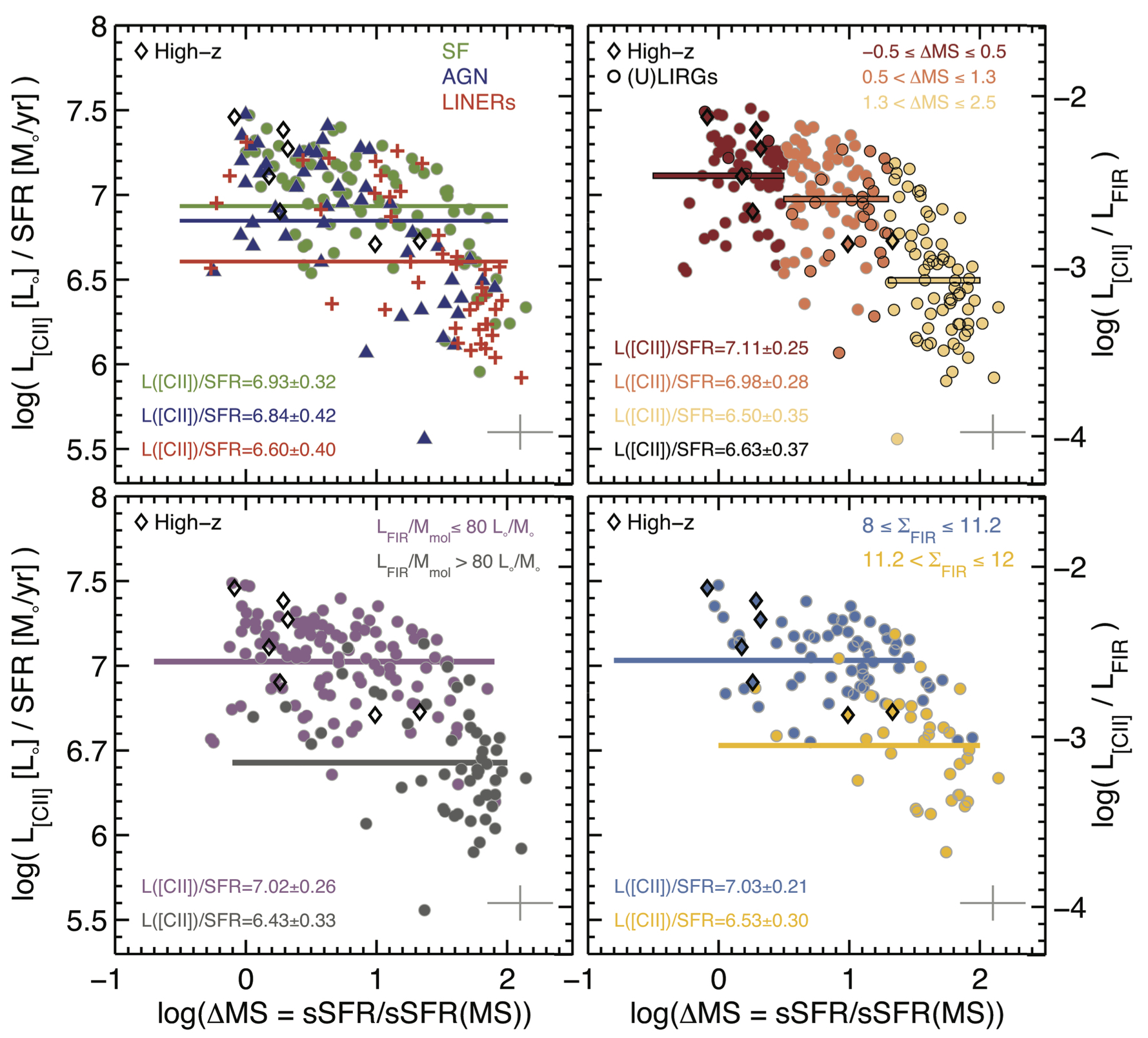}
\caption{$L_{\rm [CII]}$ - SFR and $L_{\rm [CII]}-L_{\rm FIR}$ ratios as a function of various galaxy properties. The mean value of the $L_{\rm [CII]}/{\rm SFR}$ ratio for each subgroup of galaxies is listed in the bottom-left corners of the panels and also over plotted as an horizontal color line. High-$z$ galaxies ($z>2$) are shown as diamonds. Panels: {\it (Top-left)} Nuclear optical classification (star-forming, AGN, or LINER) according to the BPT diagram. {\it (Top-right)} offset from the main-sequence. In this panel we also mark galaxies classified as (U)LIRGs with a black circle. {\it (Bottom-left)} Star formation efficiency (${\rm SFE}=L_{\rm FIR}/M_{\rm mol}$). {\it (Bottom-right)} far-infrared surface brightness \sfir. Star formation rates were calculated based on the total far-infrared using \cite{rhc_murphy11} calibration. Typical errorbars are shown in the lower-right corner.}\label{cii_sfr}
\end{center}
\end{figure*}

For the ULIRGs in our sample we only have global measurements, which implies that any impact of the AGN on the \cii/FIR ratio will be diluted by the emission from the star-forming disk. One alternative, however, to assess the influence of the AGN is to consider in the analysis what fraction of the bolometric luminosity is contributed by the nuclear activity.
This fraction is found to be, on average, $\sim30\%$ in \hii-like and LINERs ULIRGS and $\sim50-75\%$ in warm Seyfert ULIRGs \citep{rhc_veilleux09}. We observe that all the ULIRGs in our sample are characterized by very low \cii/FIR ratios ($5\times10^{-4}\lesssim{\rm \cii/FIR}\lesssim2\times10^{-3}$), and this seems to be independent of how much of the bolometric luminosity is dominated by the AGN, or how powerful is the AGN in terms of X-ray luminosity. The Kendall $\tau$ correlation coefficient of \cii/FIR versus AGN luminosity and AGN fraction are $\tau=0.06$ ($p=0.69$) and $\tau=-0.18$ ($p=0.23$), respectively. In fact, ULIRGs show \cii/FIR ratios as low as those predicted in the most extreme cases of the \cite{rhc_langer15} model, and comparable to those observed in the $z\sim1$ AGN systems that have X-ray luminosities $L_{\rm X-ray}\gtrsim10^{45}$~erg~s$^{-1}$ \citep{rhc_stacey10}. The situation is similar for the \oi\ line, as we do not observe a trend of increasing (or decreasing) \oi/FIR ratios as a function of X-ray luminosity (Kendall $\tau=0.25$ with $p=0.13$) or AGN fraction (Kendall $\tau=-0.08$ with $p=0.63$). 

In summary, we observe evidence for central suppression of the \cii/FIR ratio in the inner $\sim$500~pc of Seyfert galaxies. However, the magnitude of this drop is similar to that observed in central regions of some star-forming galaxies, or galaxies that host weak AGNs \citep[this work;][]{rhc_parkin13,rhc_rhc15,rhc_smith17}. This could be explained by the different physical conditions that dominate the nucleus and the disk regions, including for the former warmer dust temperatures, the prevalence of \oi\ as the main cooling channel, and in some cases additional contribution to the FIR emission by populations of old stars \citep[e.g.,][]{rhc_smith17}. Perhaps the only Seyfert galaxy in our sample where we find evidence for the central suppression of the \cii/FIR ratio as a direct consequence of the AGN X-ray emission is NGC~1068. Finally, in our sample of ULIRGs we do not observe any correlation between the global \cii/FIR ratio and the X-ray luminosity, or the fractional contribution of the nuclear activity to the bolometric luminosity. Thus we do not find that AGN activity plays a major role in the observed (global) \cii\ deficit of ULIRGs with $L_{\rm X-ray}\lesssim10^{44}$~erg~s$^{-1}$.

\begin{table*}[ht!]
\begin{center}
\caption{Summary of $L_{\rm [CII]}-{\rm SFR}$ and $L_{\rm [CII]}-L_{\rm FIR}$ scalings.\label{tab:scaling_LCIISFR}}
\begin{tabular}{cccc}
\hline
\hline
Class & Criteria & $log_{10}(L_{\rm [CII]}/{\rm SFR})^{a}$ & $log_{10}(L_{\rm [CII]}/L_{\rm FIR})$ \\ \hline
Star-forming & BPT Classification$^{b}$ & $6.93\pm0.32$ & $-2.62\pm0.33$ \\
AGN & BPT Classification$^{b}$  & $6.84\pm0.41$ & $-2.75\pm0.40$ \\
LINERs & BPT Classification$^{b}$  & $6.60\pm0.40$ & $-2.96\pm0.40$ \\
(U)LIRGs & $L_{\rm FIR}\geq10^{11}~L_{\odot}$ & $6.63\pm0.37$ & $-2.95 \pm0.37$ \\
\hline
Main-sequence (MS) & $1/3\leq \Delta {\rm MS}\leq3$ & $7.11\pm0.25$ & $-2.47\pm0.25$ \\
Above MS $\times3-20$ & $3< \Delta {\rm MS}\leq20$ & $6.97\pm0.28$ & $-2.61\pm0.28$ \\
MS outliers $\times20-100$ & $20< \Delta {\rm MS}\leq100$ & $6.50\pm0.34$ & $-3.08\pm0.34$ \\
\hline
Normal SFE$^{c}$ & $L_{\rm FIR}/M_{\rm mol}\leq80~L_{\odot}/M_{\odot}$ & $7.02\pm0.26$ & $-2.55\pm0.26$ \\
High SFE  & $L_{\rm FIR}/M_{\rm mol}>80~L_{\odot}/M_{\odot}$ & $6.43\pm0.33$ & $-3.14\pm0.34$ \\
\hline
Normal \sfir\  & $10^8\leq \Sigma_{\rm FIR}^{d} \leq10^{11.2}$ & $7.03\pm0.21$ & $-2.56\pm0.21$ \\
High \sfir\ & $10^{11.2} < \Sigma_{\rm FIR}\leq10^{12}$ & $6.53\pm0.30$ & $-3.05\pm0.30$ \\
\hline
\end{tabular}
\end{center}
\tablenotemark{a} Mean $\pm$ 1-$\sigma$ standard deviation. Units are $L_{\odot}/(M_{\odot}~{\rm yr}^{-1})$. SFRs are calculated using \cite{rhc_murphy11} calibration based on $L_{\rm TIR}(8-1000~\mu{\rm m})$, where  $L_{\rm TIR}(8-1000~\mu{\rm m})=1.75\times L_{\rm FIR}(42.5-122.5~\mu{\rm m})$. \\
\tablenotemark{b} Following the Baldwin,
Phillips and Terlevich (BPT) diagnostic diagram introduced in \cite{rhc_baldwin81} and \cite{rhc_veilleux87}. \\
\tablenotemark{c} SFE = Star Formation Efficiency. \\
\tablenotemark{d} Units are $L_{\odot}~{\rm kpc}^{-2}$.\\
\end{table*}

\subsection{\cii-SFR scaling relations for different galaxy types and physical conditions}\label{CIISFR_scaling}

In thermal balance, the cooling of the neutral atomic gas dominated by the \cii\ transition, traces the amount of heating by the star formation activity via the photoelectric effect on small dust grains \citep{rhc_hollenbach99}. This, combined with the high brightness of the \cii\ line, motivates the use of \cii\ as a star formation tracer \citep[e.g.,][]{rhc_boselli02,rhc_delooze14,rhc_rhc15}. However, if we assume that the FIR emission is a reliable tracer of the star formation activity \cite[e.g.,][]{rhc_kennicutt12}, then the robustness of this tracer is limited by the large variations observed in the \cii\ to FIR ratio. In order to explore the robustness of the star formation rate as a predictor of the \cii\ luminosity (or the reliability of the \cii\ transition as a star formation tracer) in this section we present scaling relations between \cii\ emission and star formation rate for different types of galaxies and physical conditions.

In order to facilitate the comparison between our sample of nearby galaxies and those at high-$z$, we present our results in the context of the ``main-sequence'' (MS) of star-forming galaxies,  i.e., the $\pm0.3$~dex scatter sequence relating star formation activity with galaxy stellar mass as a function of redshift \citep[e.g.,][]{rhc_daddi07,rhc_elbaz07,rhc_rodighiero11,rhc_whitaker12}. For this we calculate the offset from the main-sequence, $\Delta{\rm MS}$, which corresponds to the logarithm of the specific star formation rate (${\rm sSFR= SFR}/M_{*}$, where $M_{*}$ is the stellar mass) normalized by the stellar mass- and redshift-dependent center line of the main-sequence, ${\rm sSFR(MS},z,M_{*})$. For the latter we adopt the prescription proposed by \cite{rhc_speagle14}:

\begin{equation}
\begin{split}
{\rm log_{10}(sSFR(MS}, z, M_{*}))
= (-0.16-0.026t)\\ 
\times({\rm log_{10}}(M_{*})-0.025)+(2.49+0.11t)~({\rm Gyr}^{-1}), \\ \\
\end{split}
\end{equation}

\noindent where $t$ is the cosmic time in units of Gyr, and we assume a flat $\Lambda$CDM Universe with $\Omega_{m}=0.3$ and $H_{0}=70$~km~s~$^{-1}$~Mpc$^{-1}$. This prescription can be applied to galaxies in the redshift range $z=0-5$ and stellar masses in the $M_{*}=10^{9-11.8}$~$M_{\odot}$ range.

To determine the specific star formation rate of the galaxies in our sample we need stellar masses and star formation rates. We calculate stellar masses using the Two Micron All Sky Survey (2MASS) $K$-band (2.2~$\mu$m) photometry and the $M_{*}/L$ conversion from \cite{rhc_lacey08}. K-band luminosities can suffer from contamination from power-law emission from AGN heating, although we expect this contamination to be minor in Seyfert 1 and 2 sources \citep[$\lesssim30\%$ according to][]{rhc_mushotzky08}. Note, however, this may not be the true for (U)LIRGs in our sample with high AGN fractions \citep[e.g., ][]{rhc_veilleux02,rhc_veilleux06,rhc_veilleux09b}. For the systems in our sample that overlap with GOALS, we measure stellar masses that are consistent with those from \cite{rhc_howell10} (also based on 2MASS $K$-band magnitudes) and \cite{rhc_vu12} (based on SED fitting of the UV-NIR part of the spectrum). 

To calculate the star formation rates we use the calibration based on the total far-infrared luminosity ($L_{\rm TIR}(8-1000~\mu{\rm m})$) by \cite{rhc_murphy11}. We convert FIR($40-122~\mu$m) luminosities into TIR($8-1000~\mu$m) luminosities by scaling the former by a factor 1.75 \citep[see Section 5.2 in Paper I; ][]{rhc_rhc18a}. How reliable is to use the FIR emission as a star-formation rate tracer? For galaxies with $L_{\rm TIR}\gtrsim10^{11}~L_{\odot}$, TIR emission represents the most reliable SFR indicator as the un-obscured contribution of massive stars to the total SFR is small \citep[$\lesssim15\%$;][]{rhc_calzetti10}. In the intermediate luminosity range between $L_{\rm TIR}\sim10^{10}-10^{11}~L_{\odot}$, galaxies become more transparent at UV and optical wavelengths, and as a consequence the infrared emission by itself is less representative of the total star formation activity. However, as pointed out by   \cite{rhc_kennicutt12}, in this regime the effects of partial dust attenuation and dust heating by old stars roughly compensate for each other, which implies that TIR emission still trace an important fraction of the star formation activity. Finally, in galaxies with low dust content \citep[$L_{\rm TIR}\lesssim5\times10^{9}~L_{\odot}$;][]{rhc_madden13}, most of the emission produced by young, massive stars escape unabsorbed by dust, and the far infrared emission becomes a poor tracer of the star formation rate \citep[e.g.; ][]{rhc_calzetti10,rhc_delooze14}. In the sample of galaxies shown in Figures~\ref{cii_MS} and \ref{cii_sfr}, 54\% have $L_{\rm TIR}>10^{11}~L_{\odot}$, 41\% have $L_{\rm TIR}$ in the $10^{10}-10^{11}~L_{\odot}$ range, and only 5\% have TIR luminosities between $6\times10^{9}$ and $10^{10}~L_{\odot}$. In conclusion, we expect TIR emission to be a robust tracer of the star formation activity of the galaxies in our sample.

Figure~\ref{cii_MS} shows the \cii-to-FIR ratio as a function of the offset from the main-sequence of galaxies ($\Delta{\rm MS}$). We include local star-forming galaxies (circles), AGNs (squares), and high-$z$ galaxies (diamonds). For the latter we select from the literature galaxies with $L_{\rm FIR}$, $M_{*}$, SFR, sizes, and \cii\ luminosities available. These systems are: HDF850.1 \citep[$z=5.2$;][]{rhc_walter12}, HFLS3 \citep[$z=6.34$;][]{rhc_riechers13}, ALESS73.1 \citep[$z=4.8$;][]{rhc_debreuck14}, and four main-sequence, $z\sim5$ galaxies from the \cite{rhc_capak15} sample (HZ4, 6, 9 and 10). We color code galaxies according to their FIR surface brightness (when size measurements are available). 
We observe that galaxies lying within a factor of $\sim4$ of the main-sequence have \cii-to-FIR ratios in the $\sim10^{-2}-10^{-3}$ range, while outliers above the main-sequence ($\Delta{\rm MS}\gtrsim10$) have \cii-to-FIR ratios lower than $\sim3\times10^{-2}$, irrespective of whether they are classified as star-forming, AGN, or high-$z$ \citep[see also][]{rhc_gracia-carpio11,rhc_diaz-santos13,rhc_ibar15}. This trend of decreasing \cii-to-FIR ratio as a function of $\Delta{\rm MS}$ (Kendall $\tau=-0.45$ with $p<0.01$) is consistent with those observed as a function of \sratio\ and $L_{\rm FIR}/M_{\rm mol}$ \citep[see Figure~4 in][]{rhc_rhc18a}, since galaxies step along in sSFR at fixed $M_{*}$ and $z$, they have higher dust temperatures ($\propto$~\sratio), star formation efficiencies ($\propto L_{\rm FIR}/M_{\rm mol}$)\citep[e.g.,][]{rhc_genzel15,rhc_tacconi17}, and FIR surface brightnesses \citep[e.g.,][]{rhc_lutz16}.

In addition to the trend with the offset from the main-sequence, we find that in general, for a fixed value of $\Delta{\rm MS}$ the \cii-to-FIR ratio decreases with increasing \sfir. For example, for starburst galaxies with $\Delta{\rm MS}\approx30$, those with $\Sigma_{\rm FIR}\approx10^{10}$~$L_{\odot}$~kpc$^{-2}$ have \cii-to-FIR ratios about an order of magnitude higher than galaxies with $\Sigma_{\rm FIR}\approx10^{12}$~$L_{\odot}$~kpc$^{-2}$. As we explored with our {\it toy model} (see Section~\ref{toy}), an increase in \sfir\ creates enhanced FUV radiation fields, ionization parameters and neutral gas densities that manifest themselves in lower \cii\ to FIR ratios. 

We now discuss the scaling relations between \cii\ luminosity and SFR that can be used to predict the \cii\ luminosity of a galaxy if a measurement of the SFR is available, or to attempt to measure the SFR if the \cii\ luminosity is known. Figure~\ref{cii_sfr} shows the \cii\ luminosity - SFR ratio of galaxies as function of their offset from the main-sequence $\Delta{\rm MS}$. In the first panel we group galaxies according to their BPT classification, i.e., star-forming (stars), AGN (Seyfert 1 and 2; triangles), and LINERs (crosses). Star-forming and AGN galaxies follow similar distributions (a two-dimensional two-sample Kolgomorov-Smirnov test gives a p-value $p<0.01$ for the likelihood that both groups of galaxies are distributed simirlarly), reaffirming the results from \S\ref{section:AGN} that the influence of unboscured AGN on the global \cii-to-FIR ratio is small. Using the SFR as a predictor of the \cii\ luminosity of a galaxy only based on its BPT classification yields luminosities with an uncertainty factor of $2.1$ for star-forming galaxies, and 2.5 for AGNs and LINERs. 

In the second panel of Figure~\ref{cii_sfr} we divide galaxies in three groups according to their separation from the main-sequence: main-sequence galaxies ($1/3\leq\Delta{\rm MS}<3$; red circles), star-forming galaxies above the main-sequence ($3<\Delta{\rm MS}\leq20$; orange circles), and starburst outliers ($\Delta{\rm MS}>20$; golden circles). As discussed at the beginning of this section, there is a trend of decreasing $L_{\cii}/{\rm SFR}$ ratios as $\Delta{\rm MS}$ increases, which implies that starburst outliers have, on average, $L_{\cii}/{\rm SFR}$ ratios a factor of $\sim5$ lower than main-sequence galaxies. We find that using the position of a galaxy relative to the main-sequence provides a prediction for the \cii\ luminosity with an uncertainty factor of $\sim2$. 
The panel also includes galaxies classified as LIRGs (black border circles). The majority of these correspond to the group of starburst outliers and their median $L_{\cii}/{\rm SFR}$ ratio is $4.6\times10^6$~$L_{\rm \odot}/{\rm (M_{\odot}~yr^{-1})}$, a factor of $\sim3$ lower than the median ratio in main-sequence galaxies. 

The third panel of Figure~\ref{cii_sfr} shows galaxies grouped according to their \firmol\ ratio, which is a measure of the star formation efficiency in star-forming galaxies. We split the sample in two groups choosing a threshold value of \firmol~$=80$~$L_{\odot}~M_{\odot}^{-1}$; the value where we start to find a decline in the fine structure line to FIR ratio of galaxies \citep[see also ][]{rhc_gracia-carpio11}. For systems with \firmol~$\leq80$~$L_{\odot}~M_{\odot}^{-1}$ (purple circles) we observe a tight correlation (0.26~dex dispersion) over two orders of magnitude in $\Delta{\rm MS}$. This implies that if the position of the galaxy in the main-sequence plane is unknown, then information on the \firmol\ ratio can lead to predictions of the \cii\ luminosities with an uncertainty factor of $\sim2$.

Finally, the last panel in Figure~\ref{cii_sfr} show galaxies grouped according to their FIR surface brightness.  Those with \sfir\ in the $10^8-10^{11.2}$~$L_{\odot}~{\rm kpc}^{-2}$ range show the tightest correlation among all the categories previously discussed. The scatter is only 0.21 dex over two orders of magnitude in $\Delta{\rm MS}$. Galaxies with $\Sigma_{\rm FIR}>10^{11.2}$~$L_{\odot}~{\rm kpc}^{-2}$ show a correlation with a 1-$\sigma$ dispersion of 0.3~dex. The high-$z$ systems in our sample follow these trends, which suggest that the dominant physical conditions set by the compactness of the source produce similar [CII]/SFR ratios in local and high-$z$ galaxies. This also implies that the scaling relations derived based on observations of nearby galaxies --and that rely on a third parameter such as $\Delta{\rm MS}$ or \sfir-- can also be applied to high-$z$ galaxies.

One caveat is the AGN contribution to the IR luminosity that can lead to the overestimation of the SFR. This effect can be particularly important in ULIRGs \citep[e.g.,][]{rhc_veilleux09}. Note, however, that given the similar distribution of star-forming and AGN galaxies as a function of $\Delta$MS (first panel of Figure~\ref{cii_sfr}), the mean [CII]/SFR (or [CII]/FIR) ratios for the different subgroups of galaxies change by $\lesssim0.1~dex$ if we exclude AGN systems from the analysis. 

In summary, the \cii\ luminosity of a galaxy can be predicted based on its SFR (or the SFR can be calculated based on the \cii\ luminosity) with different levels of uncertainty depending which other galaxy properties are known. We find that the combination of the SFR and the FIR surface brightness produce the most robust predictions of \cii\ luminosities, with an uncertainty of only 0.21~dex if the galaxy is in the $\Sigma_{\rm FIR}=10^8-10^{11.2}$~$L_{\odot}~{\rm kpc}^{-2}$ range. This is consistent with the tight (0.21~dex) correlation between $\Sigma_{\rm [CII]}-\Sigma_{\rm SFR}$ observed in resolved regions of nearby, star-forming galaxies \citep{rhc_rhc15}. 

The second best alternative to predict \cii\ luminosities relies on the combination of the SFR with the separation of the galaxy with respect to the main-sequence, $\Delta{\rm MS}$, or the \firmol\ ratio. In these cases \cii\ luminosities can be predicted within an uncertainty of factor $\approx2$. Finally, taking as an additional parameter to the SFR the emission line classification of the galaxy --i.e., SF, AGN, or LINER-- does not contribute to a better prediction of the \cii\ luminosity. In these cases the scatter associated to the observed \cii/SFR ratio is of the order of $\sim0.35-0.4$~dex.

\subsection{Mass-metallicity relation of (U)LIRGs assessed by extinction insensitive metallicity diagnostics}\label{M-Z}

\begin{figure*}
\begin{center}
\includegraphics[scale=0.22]{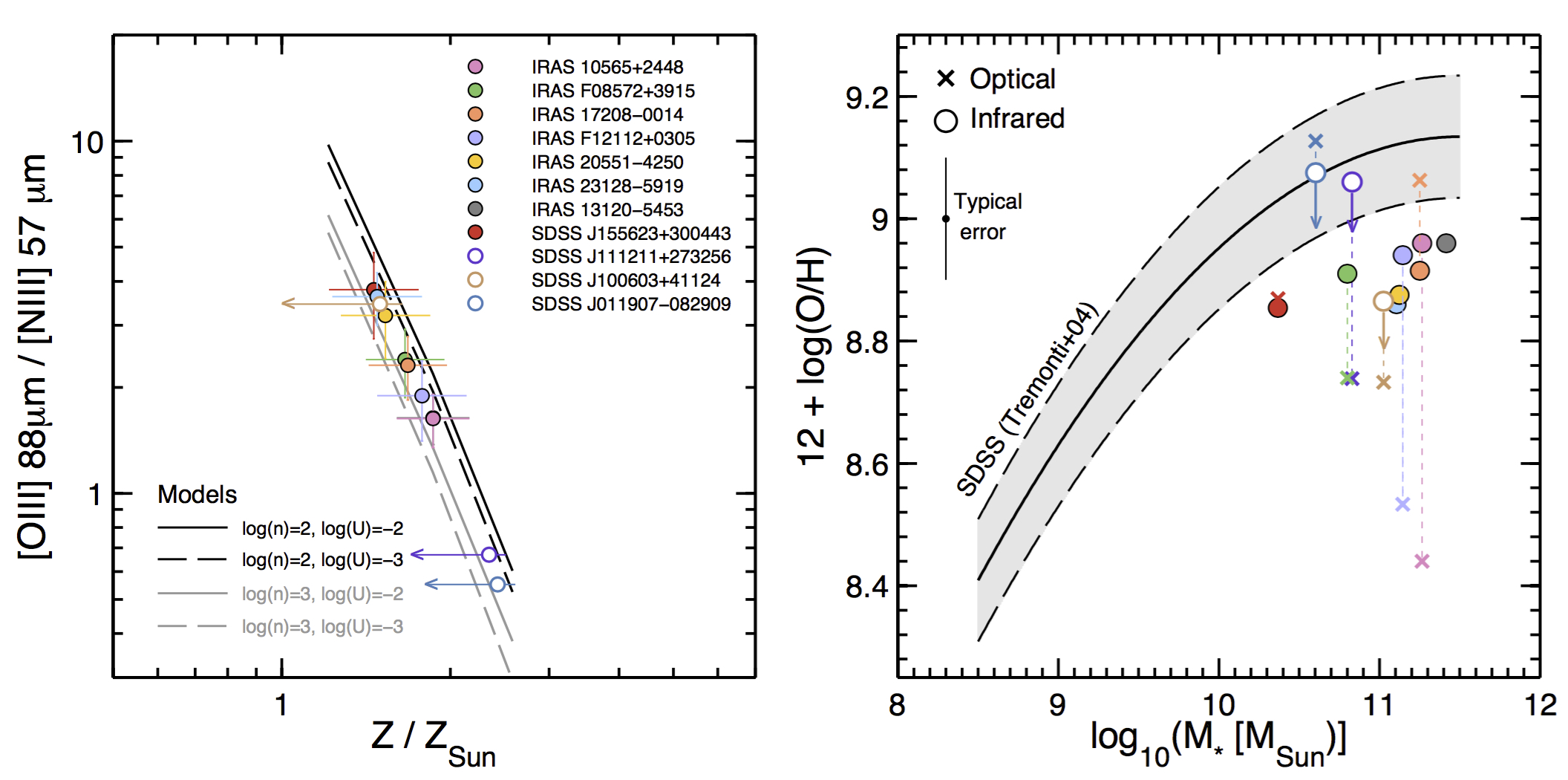}
\caption{{\it (Left)} \oiii/\niii\ emission line ratio as a function of metallicity based on the photoionization models reported in \cite{rhc_ps17} \citep[see also][]{rhc_nagao11}. The models (black and grey lines) assume ionized gas densities and ionization parameters in the 10$^{2}-10^{3}$~cm$^{-3}$ and $10^{-3}-10^{-2}$ range, respectively. The observed \oiii/\niii\ line ratio for 11 (U)LIRGs, and the corresponding abundance range expected from the model, are shown with filled circles when both lines are detected, and open circles when only the \oiii\ line is detected. {\it (Right)} Mass-metallicity relation observed in local galaxies \citep{rhc_tremonti04} and (U)LIRGs whose metallicities are measured using optical-based \citep[crosses;][]{rhc_tremonti04,rhc_rupke08,rhc_hou09} and IR-based (circles; this work) methods. All metallicity measurements are shown in the \cite{rhc_tremonti04} scale. These results confirm that (U)LIRGs tend to have lower oxygen  abundances compared to star-forming galaxies of the same stellar mass, but shows that this offset is smaller than previously thought from studies based on optical-based metallicities \citep[e.g.,][]{rhc_rupke08}.}\label{IRmetal}
\end{center}
\end{figure*}

The mass-metallicity relation is a well defined relationship between stellar mass and gas phase abundance observed in galaxies up to $z\sim3$ \citep[e.g.,][]{rhc_lequeux79,rhc_tremonti04,rhc_savaglio05,rhc_erb06,rhc_lee06,rhc_maiolino08,rhc_zahid11,rhc_ewuyts14,rhc_steidel14,rhc_ewuyts16}. The existence of this relation has been interpreted as a consequence of the interplay between star-formation, gas outflows, and gas accretion during the evolution of galaxies. While the mass-metallicity relation holds for low-metallicity and star-forming galaxies, (U)LIRGs have been found to deviate in the sense that their gas phase abundance --as inferred from optical nebular lines-- is much lower than expected from their stellar mass \citep[e.g.,][]{rhc_rupke08,rhc_kilerci14}.

The observed offset of local (U)LIRGs from the mass-metallicity relation could have at least two explanations. First, as shown by theoretical models and numerical simulations \citep[e.g.,][]{rhc_naab06,rhc_montuori10,rhc_rupke10,rhc_torrey12}, tidal forces acting in merging/interacting galaxies drive low-metallicity gas from the outskirts towards the central active star forming regions; hence the observed nuclear metallicity under-abundances and shallower metallicity gradients \citep{rhc_rupke08,rhc_kewley10,rhc_kilerci14}. Second, the low gas metallicity inferred from the optical nebular lines may not be representative of the metallicity of the heavily obscured bulk of the gas in (U)LIRGs. Two pieces of evidence that support this hypotheses are the large dust masses found in ULIRGs which are incompatible with the low metallicities inferred from the optical lines \citep[e.g,][]{rhc_santini10}, and the $\sim3\times$ solar neon abundance found in the average spectrum of 27 PAH-dominated ULIRGs by \cite{rhc_veilleux09} \citep[see also ][]{rhc_verma03}. Note however that both of these measurements are subject to uncertainties, and ideally one would want an independent, reliable, and extinction insensitive determination of the metallicity to validate one of the two possible scenarios.

Far-infrared fine structure lines originating in HII regions can work as powerful tools to measure the gas metallicity even in highly obscured star forming regions.  Based on a grid of photoionization models, \cite{rhc_nagao11}, \cite{rhc_fo16} and \cite{rhc_ps17} have shown that the flux line ratios of (\oiiii+\oiii)/\niii\ or \oiii/\niii\ are sensitive diagnostics of the gas metallicity at $Z>0.2~Z_{\odot}$. These line ratios present some residual dependence on the ionization parameter $U$ and on the ionized gas density, however, these quantities can be constrained using a combination of far-infrared \hii\ and PDR lines as we have shown in Section~\ref{cloudy}. 

In this section we use the \niii\ and \oiii\ lines to determine the extinction-insensitive metallicity of 11 (U)LIRGs. Eight of these systems have optical gas-phase metallicities available: four selected from SHINING, and the other four selected from the {\it Sloan Digital Sky Survey} (SDSS) and observed with {\it Herschel}/PACS (PI R. Maiolino; see Table~\ref{table: maiolino} for details). These systems do not show evidence for AGN activity. 

For the SDSS galaxies, stellar masses and optical metallicities were drawn from the JHU/MPA value-added galaxy catalogue\footnote{\url{http://home.strw.leidenuniv.nl/~jarle/SDSS/}} \citep{rhc_kauffmann03,rhc_tremonti04,rhc_brinchmann04}. For the SHINING galaxies IRAS F10565+2448 and IRAS F08572+3915 we use oxygen abundances in the \cite{rhc_tremonti04} calibration scale as listed in \cite{rhc_rupke08}. For IRAS 17208-0014 and IRAS F12112+0305, we draw H$\alpha$ and [NII]~6584 fluxes from \cite{rhc_moustakas06} and \cite{rhc_hou09}, and then derived oxygen abundances based on the [NII]/H$\alpha$ diagnostic in \cite{rhc_maiolino08}. We converted these metallicities to the \cite{rhc_tremonti04} scale using the conversion factor in \cite{rhc_kewley08}. 

The left panel of Figure~\ref{IRmetal} shows the measured \oiii/\niii\ line ratios (and lower-limits) as a function of metallicity based on the photoionization models reported in \cite{rhc_ps17}. The model results we include are based on a range of ionization parameters and density values typical of (U)LIRGs as estimated in Section~\ref{cloudy} \citep[see also][]{rhc_gracia-carpio11,rhc_fischer14}. The model curves shown in Figure~\ref{IRmetal} have been scaled to take into account the differences in the \cite{rhc_ps17} and \cite{rhc_tremonti04} metallicity calibrations. While the latter uses the model of \citet[][hereafter CL01]{rhc_charlot01}, the infrared method in \cite{rhc_ps17} scales the nitrogen abundance with oxygen following the fit to the observed relation in \citet[][hereafter PL14]{rhc_pilyugin14}. Both models CL01 and PL14 have parameterizations of the N/O ratio that increase as a function of O/H at a similar rate for $Z>0.25Z_{\odot}$, but with different normalization values. At a solar metallicity, the CL01 model yields a N/O abundance that is a factor of $\sim2$ lower than the N/O abundance in the PL14 calibration ($log_{10}({\rm N/O})_{\rm CL01}\approx-0.95$ versus $log_{10}({\rm N/O})_{\rm PL14}\approx-0.65$). According to the scaled version of the \cite{rhc_ps17} models, the IR-based metallicities of the (U)LIRGs in our sample are in the $1.5\lesssim Z/Z_{\odot} \lesssim2.5$ range.

How do the IR-based metallicity measurements compare to those obtained from optical nebular lines? The right panel of Figure~\ref{IRmetal} shows the comparison between both methods in the context of the observed mass-metallicity relation in local ($z\sim0.1$), star-forming galaxies from SDSS \citep{rhc_tremonti04}. Out of the five (U)LIRGs detected in both \oiii\ and \niii\ lines, we find that the two systems with the lowest optical metallicities --IRAS~F12112+0305 and IRAS~F10565+2448-- have infrared-based metallicities a factor of $\sim3$ higher. The remaining three (U)LIRGs have optical- and infrared-based metallicities that are consistent within the error bars. For IRAS~F12112+0305 and IRAS~F10565+2448 this would imply that the gas phase metallicities derived from optical and infrared methods trace different layers, in the sense that gas in the densest and dustiest star-forming regions that can only be probed by infrared lines is more enriched than the gas in less obscured regions \citep{rhc_veilleux09,rhc_santini10}. 

In summary, we find that the oxygen abundance of non-AGN (U)LIRGs derived from both optical and infrared emission line tracers tend to be lower than the metallicity of star-forming galaxies of similar stellar mass. This is in agreement with previous mass-metallicity studies of (U)LIRGs \citep{rhc_rupke08,rhc_kilerci14,rhc_ps17}. Note, however, that using infrared-based metallicities reduces the offset from the mass-metallicity relation previously found using optical line diagnostics. Among the two competing hypothesis to explain the offset from the mass metallicity relation --massive inflows of metal poor gas or the bulk of the gas being metal rich, but heavily embedded in dust-- our comparison of optical and infrared based metallicity measurements favor the former.

\section{Summary and Conclusions}\label{conclusions}

In this paper we investigate the physical mechanisms behind the the observed line deficits in galaxies. We also investigate the scaling relations between the \cii\ emission and the star formation rate, and the location of the LIRGs in the well known mass-metallicity relation of star-forming galaxies. The analysis presented here is based on the SHINING survey of galaxies \citep[Paper~I; ][]{rhc_rhc18a} which was conducted with the PACS spectrometer on board the {\it Herschel Space Observatory}.


We highlight the following points:

\begin{enumerate}
\item{{\it A Toy Model to explain the PDR lines deficit.} To explore the tight relationship observed between the \cii\ to continuum ratio and \sfir, we created a {\it toy model} that treats the ISM a the combination of two extreme scenarios. {\it Scenario 1 -- Dense PDR:} we assume that all OB stars and molecular gas clouds are closely associated. We also consider that radiation pressure in \hii\ regions acts to concentrate the gas in a spherical shell, which results in a higher density of the neutral gas confining the ionized gas than if we assume a uniform-density \hii\ region \citep{rhc_draine11}. In this scenario we find that the \cii-to-FIR ratio starts to decrease at around \sfir~$\sim3\times10^{10}$~$L_{\odot}$~kpc$^{-2}$ as at this value (1) the density of the neutral gas becomes higher than the critical density of the \cii\ line for collisional excitation by H atoms, and (2) the ionization parameter reaches a limit value of $U\approx0.01$ -- the threshold value of $U$ at which the fraction of UV photons absorbed by dust in the \hii\ region becomes important \citep{rhc_luhman03,rhc_gracia-carpio11}. This implies that the \cii\ intensity is only weakly dependent on \gof\ and \nh, while the FIR intensity remains proportional to \gof. {\it Scenario 2 -- Average ISRF:} we assume that OB associations and neutral gas clouds are randomly placed in the galactic disk. In this case we find that at  \sfir~$\sim10^{10}$~$L_{\odot}$~kpc$^{-2}$ the \cii-to-FIR ratio starts to decline as a function of \sfir\ because the \cii\ intensity becomes nearly independent of \gof --as opposed to the FIR intensity that remains proportional to this quantity--, and the photoelectric heating efficiency decreases by two orders of magnitude.

Compared to the observations, the combination of the output from both model scenarios is successful in reproducing the decline of the \cii-to-FIR ratio as a function of \sfir\ starting at about \sfir~$\approx10^{10}$~$L_{\odot}$~kpc$^{-2}$. This make sense as the real structure of the ISM is porous, therefore only a fraction of the photons produced by massive stars interact with the surrounding dense neutral gas ({\it Scenario 1}), while the rest illuminates neutral gas clouds in the galactic disk ({\it Scenario 2}). 

The threshold value of \sfir~$\approx3\times10^{10}$~$L_{\odot}$~kpc$^{-2}$ at which our model predicts a sharper decline in the \cii/FIR ratio is remarkably similar to the \sfir\ value that differentiate galaxies forming star in normal or starburst mode \citep[e.g.,][]{rhc_elbaz11}, and PDRs from having constant or increasing \gof/\nh\ ratios \citep{rhc_diaz-santos17}. This indicates that the properties of the PDR/\hii\ region complexes in these two group of galaxies are significantly different, which leads to the observed differences in their global \cii/FIR ratios.

Our {\it toy model} is also successful reproducing the trends observed for the \oi\ and \oii\ lines.}\\

\item{{\it Cloudy modeling.} In addition to the {\it toy model} calculations, we use the {\it Cloudy}-based models described in \cite{rhc_fischer14} and \cite{rhc_abel09} to study the trends observed in the line to continuum ratios of galaxies as a function of the ionization parameter $U$ and the hydrogen column density \NH, for a simple shell geometry. We find that the observed decrease in the (\cii+\oi+\oii)/FIR, \nii/FIR, and \niii/\nii\ ratios as the FIR surface brightness increases from \sfir$\sim10^{10}$ to $10^{12}$~$L_{\odot}$~kpc$^{-2}$ can be explained by increasing \NH\ from $\sim10^{23.5}$ to $\sim10^{24}$~cm$^{-2}$ and $U$ from $\sim10^{-3}$ to $\sim10^{-2}$. These results are consistent with previous studies \citep[e.g.,][]{rhc_gracia-carpio11,rhc_fischer14}, and also the interpretation from our {\it toy model} in the scenario where OB stars are closely associated to molecular clouds.
}

\item{{\it AGN impact on the \cii\ and \ois\ line emission.} In Seyfert galaxies that can be spatially resolved by {\it Herschel}/PACS we find that central regions have \cii-to-FIR ratios a factor of $\sim2-6$ lower than the median ratio in the disk. In contrast, we find central \oi-to-FIR ratios that are comparable to or even higher than the median disk value. We also find very strong \oi, \oii\ and \oiii\ emission in the central spaxel of compact \hii\ Seyfert galaxies. This is probably a consequence of the denser gas and the harder radiation fields to which the ISM is exposed in the central AGN regions, which favors the cooling of the neutral gas via the \oi\ line emission. Models of \cite{rhc_abel09} and \cite{rhc_fischer14} also show that this ratio increases for high $U$ and high density. One additional factor that can contribute to the \cii\ deficit observed in AGN is the change in the C$^{++}$/C$^{+}$ balance due to the hardness of the AGN radiation field \citep{rhc_langer15}. The only case where we find this effect could have a significant impact on the \cii-to-FIR ratio is in the central region of Seyfert galaxy NGC~1068.

In contrast, we find that AGN activity does not play a major role in setting the global-scale \cii\ to FIR ratio in (U)LIRGs.}

\item{{\it Scaling relations.} One of the goals of this paper is to provide a reference sample that can be used to analyze local and high-$z$ infrared line observations of galaxies. In addition to the tables with line and continuum fluxes for the full SHINING sample listed in Paper~I \citep{rhc_rhc18a}, here we present scaling relations for the \cii-to-SFR ratio as a function of galaxy type (\hii, AGN, LINERs, and LIRGs), \firmol, \sfir, and separation from the main-sequence of galaxies ($\Delta{\rm MS}$). These can be useful for those interested in planning \cii\ observations of local or high-$z$ galaxies, or those who want to use the \cii\ line as a SFR tracer. 
We conclude that the most reliable method to predict \cii\ fluxes --or to measure SFRs from \cii\ fluxes-- is when information on the infrared size of the source is available. This is likely due to the fact that normalized quantities such as \sfir\ or $\Sigma_{\rm SFR}$ are more representative of \gof\ (the local FUV radiation field intensity), one of the main parameters controlling the \cii\ relative line intensity in the neutral ISM \citep[see also ][]{rhc_delooze14,rhc_rhc15,rhc_smith17}.}

\item{{\it Extinction-insensitive metallicity diagnostics of (U)LIRGs.}  We use the \oiii/\niii\ emission line ratio, in combination with the models by \cite{rhc_ps17} \citep[see also][]{rhc_nagao11}, to determine the oxygen phase-abundance of eight (U)LIRGs that, according to their optical-based metallicities, fall below the local mass-metallicity relation \citep{rhc_tremonti04}. We find that the (U)LIRGs in our sample have infrared-based metallicities in the $1.5\lesssim Z/Z_{\odot} \lesssim2.5$ range. For two of the systems the infrared-based metallicities are a factor of $\sim3$ higher than the optical-based measurements. We confirm that (U)LIRGs lie below the observed local mass-metallicity relation as previously reported based on optical recombination line studies \citep[e.g.,][]{rhc_rupke08,rhc_kilerci14}.  These results are consistent with a scenario where the oxygen under-abundance observed in (U)LIRGs is due to merger-driven massive inflows of metal poor gas from the outskirts of the disk to central regions \citep{rhc_naab06,rhc_rupke08,rhc_kewley10}, but we cannot rule out: (1) the effects of extinction of the most enriched regions even in the FIR,  and (2) that line ratios of two lines that are observed in deficit are more sensitive to the emission from the non-deficit regions \citep{rhc_fischer14}, i.e. the outer regions of the system.}

\end{enumerate}

\begin{acknowledgements}
We thank the referee for helpful and constructive comments that improved the paper. We thank Mark Wolfire, Natascha F{\"o}rster Schreiber, Shmuel Bialy, and Taro Shimizu for helpful discussions and comments. RHC would like to thank the support and encouragement from Fares Bravo Garrido and dedicates this paper with love to Fares and Olivia. Basic research in IR astronomy at NRL is funded by the US ONR.  JF also acknowledges support from the NHSC. E.GA is a Research Associate at the Harvard-Smithsonian Center for Astrophysics, and thanks the Spanish Ministerio de Econom\'{\i}a y Competitividad for support under projects FIS2012-39162-C06-01 and  ESP2015-65597-C4-1-R, and NASA grant ADAP NNX15AE56G. RM acknowledges the ERC Advanced Grant 695671 ``QUENCH'' and support from the Science and Technology Facilities Council (STFC). The Herschel spacecraft was designed, built, tested, and launched under a contract to ESA managed by the Herschel/Planck Project team by an industrial consortium under the overall responsibility of the prime contractor Thales Alenia Space (Cannes), and including Astrium (Friedrichshafen) responsible for the payload module and for system testing at spacecraft level, Thales Alenia Space (Turin) responsible for the service module, and Astrium (Toulouse) responsible for the telescope, with in excess of a hundred subcontractors. PACS has been developed by a consortium of institutes led by MPE (Germany) and including UVIE (Austria); KU Leuven, CSL, IMEC (Belgium); CEA, LAM (France); MPIA (Germany); INAF-IFSI/OAA/OAP/OAT, LENS, SISSA (Italy); IAC (Spain). This development has been supported by the funding agencies BMVIT (Austria), ESA-PRODEX (Belgium), CEA/CNES (France), DLR (Germany), ASI/INAF (Italy), and CICYT/MCYT (Spain). HCSS / HSpot / HIPE is a joint development (are joint developments) by the Herschel Science Ground Segment Consortium, consisting of ESA, the NASA Herschel Science Center, and the HIFI, PACS and SPIRE consortia. This publication makes use of data products from the Sloan Digital Sky Survey (SDSS).  Funding for the Sloan Digital Sky Survey has been provided by the Alfred P. Sloan Foundation, the Participating Institutions, the National Aeronautics and Space Administration, the National Science Foundation, the U.S. Department of Energy, the Japanese Monbukagakusho, and the Max Planck Society. This research has also made use of the NASA/IPAC Extragalactic Database (NED) which is operated by the Jet Propulsion Laboratory, California Institute of Technology, under contract with the National Aeronautics and Space Administration.
\end{acknowledgements}
 
\software{\textsc{CLOUDY} \citep{rhc_ferland13}, HIPE \citep[v13.0;][]{rhc_ott10}} 
\facility{Herschel \citep{rhc_pilbratt10}}


\appendix

\section{\niii\ and \oiii\ line observations in (U)LIRGs selected from SDSS}

\begin{table*}
\begin{center}
\caption{(U)LIRGs selected from SDSS \label{table: maiolino}}
\begin{tabular}{cccccc}
\hline \hline
Source & log($L_{\rm FIR}$) & log($M_{*}$) & 12+log(O/H) & \niii\ & \oiii\ \\
 & $L_{\odot}$ & $M_{\odot}$ & & W~m$^{-2}$& W~m$^{-2}$ \\
\hline
SDSS J155623+300443 & 11.71 & 10.36 & 8.87 & 2.41E-17$\pm$5.3E-18 & 9.14E-17$\pm$6.5E-18 \\
SDSS J011907-082909 & 12.09 & 10.60 & 9.13 & $<$5.76E-17 & 3.17E-17$\pm$2.36E-18 \\
SDSS J111211+273256 & 12.15 & 10.83 & 8.74 & $<$5.36e-17 & 3.58E-17$\pm$2.4E-18 \\
SDSS J100603+411224 & 12.49 & 11.02 & 8.73 & $<$1.28e-17 & 4.41E-17$\pm$7E-18 \\
\hline
\end{tabular}
\end{center}
\tablecomments{Stellar masses and metallicities drawn from the SDSS JHU/MPA value added catalogue.}
\end{table*}

\bibliography{references.bib}

\end{document}